\documentclass[12pt]{article}
\usepackage{amsmath}
\usepackage{amssymb}
\usepackage{graphics}

\renewcommand{\overline}[1]{\bar{#1}}
\makeatletter\@addtoreset{equation}{section}
\makeatother
\usepackage[dvips]{color} 
\begin{document}
\begin{titlepage}
\begin{flushright}
TIT/HEP-606\\
KIAS-P10027\\
September 2010
\end{flushright}
\vspace{0.5cm}
\begin{center}
{\Large \bf
$\mathcal{N}=2$ Instanton Effective Action in $\Omega$-background and 
D3/D($-1$)-brane System in R-R Background}
\lineskip .75em
\vskip1.0cm
{\large Katsushi Ito${}^{1}$, Hiroaki Nakajima${}^{2,3}$, Takuya Saka${}^{1}$ and Shin
Sasaki${}^{1}$ }
\vskip 2.5em
${}^{1}$ {\normalsize\it Department of Physics\\
Tokyo Institute of Technology\\
Tokyo, 152-8551, Japan} \vskip 1.5em
${}^{2}$ {\normalsize\it School of Physics\\
Korea Institute for Advanced Study \\
Seoul, 130-722, Korea} \vskip 1.5em
${}^{3}$ {\normalsize\it Department of Physics\\
Kyungpook National University \\
Taegu, 702-701, Korea}
\vskip 3.5em
\end{center}
\begin{abstract}
We study the relation between
the ADHM construction of instantons in the $\Omega$-background and the
fractional D3/D$(-1)$-branes at the orbifold singularity of
$\mathbb{C}\times\mathbb{C}^2/\mathbb{Z}_2$ in Ramond-Ramond (R-R) 3-form field strength
background. 
We calculate disk amplitudes of
open strings connecting the D3/D$(-1)$-branes in certain R-R background to 
obtain the D$(-1)$-brane effective action 
deformed by the R-R background. 
We show that
the deformed D$(-1)$-brane effective action agrees with the instanton effective 
action in the $\Omega$-background.
\end{abstract}
\end{titlepage}

\baselineskip=0.7cm
\tableofcontents
\section{Introduction}
The $\Omega$-background deformation provides a useful method to calculate
the instanton partition function of $\mathcal{N}=2$ super Yang-Mills
Theory
\cite{Ne}.
Using the ADHM construction of instantons (for a review, see
\cite{DoHoKhMa}), 
the path integral around an instanton solution reduces to the 
integral of the instanton effective action over the moduli space, 
which is called the instanton partition function.
The integral, however,  diverges due to
non-compactness and singularities of the moduli space,
but is regularized by deforming the effective
action in the $\Omega$-background 
\cite{MoNeSh, Ne, LoMaNe}
 and introducing spacetime 
noncommutativity
(or the Fayet-Iliopoulos term).
The $\Omega$-background is defined by dimensional reduction of 
six-dimensional metric with actions of two commuting vector fields 
$\Omega_m=\Omega_{mn}x^n$ and 
$\bar{\Omega}_m=\bar{\Omega}_{mn}x^n$ and the R-symmetry gauge field
Wlison line.
Here $\Omega_{mn}$ and  $\bar{\Omega}_{mn}$
are constant anti-symmetric matrices.
The regularized instanton partition function 
can be written in the form of the equivariant
integral by using the equivariant 
BRST operator associated with the vector field $\Omega_m$, 
and is evaluated explicitly via the localization formula, where
only the fixed points of 
the BRST operator
contribute to the integral \cite{Ne}.
We then obtain
Nekrasov's formula of the 
instanton partition function, which 
includes not only the prepotential of ${\cal N}=2$ gauge theory but also
the graviphoton correction to the prepotential.

In order to apply Nekrasov's formula to various gauge theories, it would be
important to realize the $\Omega$-background in superstring
theory.
One way to study the graviphoton correction in the framework of superstrings
 is to calculate the higher-genus partition 
functions of closed topological strings embedded in type II superstrings in the
self-dual graviphoton background 
\cite{AnGaNaTa, BeCeOoVa, IqKa, EgKa}.
It is known that the self-dual graviphoton
background corresponds to 
the self-dual $\Omega$-background where $\Omega_{mn}$ is self-dual.
For non-self-dual case,
it is necessary to introduce further refinement of the topological strings
\cite{Iqbal:2007ii, Taki:2007dh, Awata:2008ed, Huang:2010kf}
and backgrounds associated with
vector multiplets \cite{Antoniadis:2010iq}.

On the other hand, in \cite{BiFrFuLe} the authors studied a fractional 
D3/D$(-1)$-brane system in the $\mathbb{C}\times\mathbb{C}^2/\mathbb{Z}_2$ orbifold and 
in a Ramond-Ramond (R-R) closed string background.
Massless states of open strings with at least one end on a D$(-1)$-brane 
describe the ADHM moduli and auxiliary variables associated with the ADHM
constraints \cite{Douglas}.
The instanton effective action is reproduced by calculating disk 
amplitudes among the massless states.
The R-R background induces new interaction terms among them 
and deforms the instanton effective action \cite{Billo:2008sp}.
For the R-R background with self-dual 3-form field strength,
they showed that 
the deformed effective action agrees with the instanton effective action of
${\cal N}=2$ super Yang-Mills theory in the self-dual
$\Omega$-background, where both $\Omega_{mn}$ and $\bar{\Omega}_{mn}$ are 
self-dual.
Promoting the constant R-R field to a ${\cal N}=2$ chiral Weyl
multiplet,
they argued the relation between disk amplitude contributions to
the instanton partition function and the topological string
amplitudes.
In a previous paper \cite{ItNaSa2}, three of the present authors
studied the ADHM construction of instantons in $\mathcal{N}=2$ super
Yang-Mills theory in the self-dual $\Omega$-background, generalizing the 
analysis in \cite{BiFrFuLe} for $\bar{\Omega}_{mn}=0$ case.
We found
that the instanton effective action agrees with the D$(-1)$-brane effective action 
in the self-dual R-R 3-form background. 
It was further generalized to the ${\cal N}=4$ case whose deformation 
is obtained by dimensional reduction from ten-dimensional $\Omega$-background
\cite{ItNaSaSa}. 

In this paper, we will study the instanton effective action in 
general (non-(anti-)self-dual) 
$\Omega$-background 
and its relations to the D3/D$(-1)$-brane system in certain R-R 3-form
background.
Since we do not see any explicit derivation of the instanton
effective action based on the ADHM construction in the general
$\Omega$-background in previous literatures,
we will work out in detail.
The instanton effective action is shown to be 
the same as the one in \cite{Ne}
and be exact with respect to the
twisted supersymmetry which is deformed in the
$\Omega$-background.
We will then introduce an R-R 3-form background 
and calculate disk amplitudes with
an insertion of the closed string R-R vertex operator.
We will show that the deformed D$(-1)$-brane effective action agrees with the
instanton effective action in the $\Omega$-background.

The organization of this paper is as follows.
In section 2, 
we introduce four-dimensional $\mathcal{N}=2$ super Yang-Mills theory in 
the $\Omega$-background and discuss the deformed supersymmetry.
In section 3, 
we study the ADHM construction of instantons.
We compute the instanton effective action in the $\Omega$-background and 
show that the action is exact under the deformed supersymmetry.
In section 4, 
we investigate general R-R background and calculate 
the disk amplitudes in the presence of the R-R 3-form background.
Based on the amplitudes, we construct the effective action of 
D$(-1)$-branes in the fractional D3/D$(-1)$-brane system.
Section 5
is devoted to conclusions and discussion.

\section{$\mathcal{N}=2$ super Yang-Mills theory in $\Omega$-background}
In this section we discuss deformation of 
$\mathcal{N}=2$ $U(N)$ super Yang-Mills theory 
in the $\Omega$-background \cite{Ne, NeOk}. 
We will define the theory in spacetime with Euclidean signature.
$\mathcal{N}=2$ super Yang-Mills theory contains 
a gauge field $A_{m}$ ($m=1,2,3,4$), 
Weyl fermions $\Lambda^{I}_{\alpha}$, $\bar{\Lambda}^{I}_{\dot{\alpha}}$, 
and complex scalars $\varphi$, $\bar{\varphi}$. They belong to 
the adjoint representation of $U(N)$ gauge group. 
$\alpha,\ \dot{\alpha}=1,2$ denotes the spinor indices of the Lorentz group 
$SO(4)=SU(2)_{L}\times SU(2)_{R}$. 
$I=1,2$ is the index of $SU(2)_{I}$ R-symmetry. 
These $SU(2)$ indices are raised and lowered 
by the antisymmetric $\epsilon$-symbol 
with $\epsilon^{12}=-\epsilon_{12}=1$. 
We expand the fields with $U(N)$ 
basis $T^{u}$ $(u=1,2,\ldots,N^{2})$ normalized by 
$\mathrm{Tr}(T^{u}T^{v})=\kappa\delta^{uv}$ 
with a certain constant $\kappa$. 
The Lagrangian of the theory is given by 
\begin{align}
\mathcal{L}_{0}
&=
\frac{1}{\kappa}\textrm{Tr}\biggl[
\frac{1}{4}F_{mn}F^{mn}
-\frac{i\theta g^{2}}{32\pi^{2}}F_{mn}\tilde{F}^{mn}
+\Lambda^{I}\sigma^{m}D_{m}\bar{\Lambda}_{I}
+(D_{m}\varphi)D^{m}\bar{\varphi}
\notag\\
&\qquad\qquad{}
-i\frac{g}{\sqrt{2}}\Lambda^{I}[\bar{\varphi},\Lambda_{I}]
+i\frac{g}{\sqrt{2}}\bar{\Lambda}_{I}[\varphi,\bar{\Lambda}^{I}]
+\frac{g^{2}}{2}[\varphi,\bar{\varphi}]^{2}\biggr],
\label{undef}
\end{align}
where $F_{mn}=\partial_{m}A_{n}-\partial_{n}A_{m}+ig[A_{m},A_{n}]$ 
is the gauge field strength, 
$g$ is the gauge coupling constant and $D_{m}=\partial_{m}+ig[A_{m},\ast]$ 
is the gauge covariant derivative. 
We also define the Dirac matrices 
$\sigma_{m}=(i\tau^1,i\tau^2,i\tau^3,1)$ and
$\bar{\sigma}_{m}=(-i\tau^1,-i\tau^2,-i\tau^3,1)$,
where $\tau^{c}$ ($c=1,2,3$) are the Pauli matrices. 
$\theta$ is the theta-angle and 
$\tilde{F}_{mn}=\frac{1}{2}\epsilon_{mnpq}F^{pq}$ is the dual of $F_{mn}$. 
The projection of the field strength into the (anti-)self-dual part 
is given by 
$F_{mn}^{\pm}=\frac{1}{2}(F_{mn}\pm\tilde{F}_{mn})$. 
The second term of \eqref{undef} is topological and the 
instanton number is defined by 
\begin{gather}
k=\int d^{4}x\,\frac{1}{\kappa}\textrm{Tr}\biggl[
\frac{g^{2}}{32\pi^{2}}F_{mn}\tilde{F}^{mn}
\biggr]. 
\end{gather}
We note that the Lagrangian 
\eqref{undef} is obtained 
by dimensional reduction of $\mathcal{N}=1$ super Yang-Mills theory in 
six-dimensional flat spacetime to four dimensions. 

The $\Omega$-background \cite{MoNeSh, LoNeSh, Ne, LoMaNe, NeOk} 
is defined by 
a nontrivial fibration of $\mathbb{R}^{4}$ over $T^{2}$. 
We denote the complex coordinates of a two-dimensional torus $T^{2}$ by 
$(z, \bar{z})$, 
The six-dimensional metric is given by 
\begin{gather}
ds_{\mathrm{6D}}^{2}
=
2d\bar{z}dz+(dx^{m}+\Omega^{m}d\bar{z}+\bar{\Omega}^{m}dz)^{2}, 
\label{metric}
\end{gather}
where $\Omega^{m}=\Omega^{mn}x_{n}$, 
$\bar{\Omega}^{m}=\bar{\Omega}^{mn}x_{n}$. 
$\Omega^{mn}$ and $\bar{\Omega}^{mn}$ are 
constant antisymmetric matrices. 
We require that the metric \eqref{metric} is flat. 
This leads to the condition that 
$\Omega^{mn}$ and $\bar{\Omega}^{mn}$ commute with each other: 
\begin{gather}
\Omega^{mn}\bar{\Omega}_{np}
-\bar{\Omega}^{mn}\Omega_{np}
=0.
\label{commutation}
\end{gather}
{}From this condition, 
$\Omega^{mn}$ and $\bar{\Omega}^{mn}$ 
can be taken of the form 
\begin{gather}
\Omega^{mn}=\frac{1}{2\sqrt{2}}
\begin{pmatrix}
0 & i\epsilon_{1} & 0 & 0  \\
-i\epsilon_{1} & 0 & 0 & 0 \\
0 & 0 & 0 & -i\epsilon_{2} \\
0 & 0 & i\epsilon_{2} & 0  \\
\end{pmatrix}, 
\quad 
\bar{\Omega}^{mn}=\frac{1}{2\sqrt{2}}
\begin{pmatrix}
0 & -i\bar{\epsilon}_{1} & 0 & 0  \\
i\bar{\epsilon}_{1} & 0 & 0 & 0 \\
0 & 0 & 0 & i\bar{\epsilon}_{2} \\
0 & 0 & -i\bar{\epsilon}_{2} & 0  \\
\end{pmatrix},
\label{omega}
\end{gather}
where $\epsilon_{1}$, $\epsilon_{2}$ are complex parameters. 
$\bar{\epsilon}_{1}$, $\bar{\epsilon}_{2}$ are their complex conjugate 
respectively. 
We also introduce the R-symmetry Wilson line by gauging 
$SU(2)_{I}$ R-symmetry as 
\begin{align}
\boldsymbol{A}^{I}{}_{J}
=\mathcal{A}^{I}{}_{J}d\bar{z}+\bar{\mathcal{A}}^{I}{}_{J}dz,
\label{wilson}
\end{align}
where we consider the case such that the R-symmetry gauge fields
$\mathcal{A}^{I}{}_{J}$, $\bar{\mathcal{A}}^{I}{}_{J}$ are constant. 

Four-dimensional 
$\mathcal{N}=2$ super Yang-Mills theory in $\Omega$-background 
is obtained by dimensional reduction of six-dimensional 
$\mathcal{N}=1$ super Yang-Mills theory with the metric \eqref{metric} 
and the R-symmetry Wilson line \eqref{wilson}. 
The Lagrangian is given by 
\begin{align}
\mathcal{L}^{}_{\Omega}
&=
\frac{1}{\kappa}\textrm{Tr}\biggl[
\frac{1}{4}F_{mn}F^{mn}
-\frac{i\theta g^{2}}{32\pi^{2}}F_{mn}\tilde{F}^{mn}
+(D_{m}\varphi-gF_{mn}\Omega^{n})
(D^{m}\bar{\varphi}-gF^{mp}\bar{\Omega}_{p})
\notag\\
&\qquad\qquad{}
+\Lambda^{I}\sigma^{m}D_{m}\bar{\Lambda}_{I}
-\frac{i}{\sqrt{2}}g\Lambda^{I}[\bar{\varphi},\Lambda_{I}]
+\frac{i}{\sqrt{2}}g\bar{\Lambda}_{I}[\varphi,\bar{\Lambda}^{I}]
\notag\\
&\qquad\qquad{}
+\frac{1}{\sqrt{2}}g\bar{\Omega}^{m}\Lambda^{I}
D_{m}\Lambda_{I}
-\frac{1}{2\sqrt{2}}g\bar{\Omega}^{mn}\Lambda^{I}
\sigma_{mn}\Lambda_{I}
\notag\\
&\qquad\qquad{}
-\frac{1}{\sqrt{2}}g\Omega^{m}\bar{\Lambda}_{I}
D_{m}\bar{\Lambda}^{I}
+\frac{1}{2\sqrt{2}}g\Omega^{mn}\bar{\Lambda}_{I}
\bar{\sigma}_{mn}\bar{\Lambda}^{I}
\notag\\
&\qquad\qquad{}
+\frac{g^{2}}{2}\Bigl([\varphi,\bar{\varphi}]
+i\Omega^{m}D_{m}\bar{\varphi}-i\bar{\Omega}^{m}D_{m}\varphi
+ig\bar{\Omega}^{m}\Omega^{n}F_{mn}
\Bigr)^{2}
\notag\\
&\qquad\qquad{}
-\frac{1}{\sqrt{2}}g\bar{\mathcal{A}}^{J}{}_{I}
\Lambda^{I}\Lambda_{J}
-\frac{1}{\sqrt{2}}g\mathcal{A}^{J}{}_{I}
\bar{\Lambda}^{I}\bar{\Lambda}_{J}
\biggr],
\label{omega_lag}
\end{align}
where $\sigma^{mn}$ and $\bar{\sigma}^{mn}$ are the Lorentz generators 
defined by 
\begin{gather}
(\sigma^{mn})_{\alpha}{}^{\beta}=
\frac{1}{4}\bigl(
\sigma^{m}_{\alpha\dot{\alpha}}\bar{\sigma}^{n\dot{\alpha}\beta}
-\sigma^{n}_{\alpha\dot{\alpha}}\bar{\sigma}^{m\dot{\alpha}\beta}
\bigr),
\quad 
(\bar{\sigma}^{mn})^{\dot{\alpha}}{}_{\dot{\beta}}=
\frac{1}{4}\bigl(
\bar{\sigma}^{m\dot{\alpha}\alpha}\sigma^{n}_{\alpha\dot{\beta}}
-\bar{\sigma}^{n\dot{\alpha}\alpha}\sigma^{m}_{\alpha\dot{\beta}}
\bigr).
\end{gather}
Hereafter we regard $\Omega^{mn}$ and $\bar{\Omega}^{mn}$ as
independent parameters each other rather than the complex conjugate. 
The case of self-dual $\Omega^{mn}$, $\bar{\Omega}^{mn}$ 
without $\mathcal{A}^{I}{}_{J}$, $\bar{\mathcal{A}}^{I}{}_{J}$ 
was discussed in our previous paper \cite{ItNaSa2}. 
In this case, the Lagrangian \eqref{omega_lag} preserves 
a half of supersymmetries associated 
with the anti-chiral spinor parameter%
\footnote{
In the case of anti-self-dual $\Omega^{mn}$, $\bar{\Omega}^{mn}$ 
without the Wilson line gauge fields, 
the Lagrangian \eqref{omega_lag} preserves 
the other half of supersymmetries associated 
with the chiral spinor parameter. 
}
$\bar{\xi}^{I}_{\dot{\alpha}}$ 
since the supersymmetry variation of \eqref{omega_lag} 
with vanishing 
$\mathcal{A}^{I}{}_{J}$ and $\bar{\mathcal{A}}^{I}{}_{J}$ is 
written as 
\begin{align}
&\left.\delta^{\bar{\xi}}\mathcal{L}^{}_{\Omega}\right|
_{\mathcal{A},\bar{\mathcal{A}}=0}
\notag\\
&=
\frac{1}{\kappa}\textrm{Tr}\biggl[
-\frac{g}{\sqrt{2}}\Omega^{-mn}F_{mn}^{-}
\bar{\xi}^{I}\bar{\Lambda}_{I}
+\sqrt{2}gF_{mn}^{-}\Omega^{-mp}\bar{\xi}^{I}
\bar{\sigma}^{np}\bar{\Lambda}_{I}
-2g\bar{\Omega}^{-mn}\bar{\xi}^{I}\bar{\sigma}_{n}\Lambda_{I}
(D_{m}\varphi-gF_{mp}\Omega^{p})
\notag\\
&\qquad\quad{}
+\frac{i}{\sqrt{2}}g^{2}\Omega^{-mn}
\bar{\xi}^{I}\bar{\sigma}_{mn}\bar{\Lambda}_{I}
\Bigl([\varphi,\bar{\varphi}]
+i\Omega^{p}D_{p}\bar{\varphi}-i\bar{\Omega}^{p}D_{p}\varphi
+ig\bar{\Omega}^{p}\Omega^{q}F_{pq}
\Bigr)
\biggr],
\label{eq92}
\end{align}
and \eqref{eq92} vanishes for self-dual $\Omega^{mn}$ and $\bar{\Omega}^{mn}$. 
When $\Omega^{mn}$, $\bar{\Omega}^{mn}$ are not (anti-)self-dual, 
the Lagrangian \eqref{omega_lag} does not have supersymmetry in general. 
However 
if we identify the Wilson line gauge fields 
with $\Omega^{mn}$ and $\bar{\Omega}^{mn}$ such as
\begin{gather}
\mathcal{A}^{I}{}_{J}
=-\frac{1}{2}\Omega_{mn}(\bar{\sigma}^{mn})^{I}{}_{J},\quad 
\bar{\mathcal{A}}^{I}{}_{J}
=-\frac{1}{2}\bar{\Omega}_{mn}(\bar{\sigma}^{mn})^{I}{}_{J}, 
\label{condition1}
\end{gather}
then we have one anti-chiral supersymmetry%
\footnote{
If we replace $\bar{\sigma}^{mn}$ in \eqref{condition1} with $\sigma^{mn}$, 
we have one chiral supersymmetry. 
}
\cite{NeOk} associated with the parameter 
$\bar{\xi}=\delta_{I}^{\dot{\alpha}}\bar{\xi}^{I}_{\dot{\alpha}}$. 
Namely, in this case, \eqref{eq92} is canceled by the contribution from the 
R-symmetry Wilson line. 
This supersymmetry 
plays a role of topological BRST symmetry. It is convenient to write down
this symmetry using the topological twist \cite{Wi}, which is obtained by
identifying the $SU(2)_I$ R-symmetry indices $I$ with the $SU(2)_R$ 
spinor indices $\dot{\alpha}$. 
We define the twisted fields by 
\begin{gather}
\Lambda_{m}=\bar{\sigma}_{m}^{I\alpha}\Lambda_{\alpha I}, 
\quad 
\bar{\Lambda}=\delta^{\dot{\alpha}}_{I}\bar{\Lambda}_{\dot{\alpha}}^{I}, 
\quad 
\bar{\Lambda}_{mn}=-(\bar{\sigma}_{mn})^{\dot{\alpha}}{}_{I}
\bar{\Lambda}_{\dot{\alpha}}^{I}. 
\end{gather}
The supersymmetry is deformed by 
$\Omega_{mn}$ and $\bar{\Omega}_{mn}$.
We denote this supercharge as $\bar{Q}_{\Omega}$, 
which acts on the fields as 
\begin{align}
\bar{Q}^{}_{\Omega}A_{m}
&=
\Lambda_{m},
\notag\\
\bar{Q}^{}_{\Omega}\Lambda_{m}
&=
-2\sqrt{2}(D_{m}\varphi-gF_{mn}\Omega^{n}),
\notag\\
\bar{Q}^{}_{\Omega}\bar{\Lambda}
&=
-2ig\bigl([\varphi,\bar{\varphi}]
+i\Omega^{m}D_{m}\bar{\varphi}-i\bar{\Omega}^{m}D_{m}\varphi
+ig\bar{\Omega}^{m}\Omega^{n}F_{mn}
\bigr),
\notag\\
\bar{Q}^{}_{\Omega}\bar{\Lambda}_{mn}
&=
2H_{mn},
\notag\\
\bar{Q}^{}_{\Omega}\varphi
&=
g\Omega^{m}\Lambda_{m},
\notag\\
\bar{Q}^{}_{\Omega}\bar{\varphi}
&=
-\sqrt{2}\bar{\Lambda}
+g\bar{\Omega}^{m}\Lambda_{m},
\notag\\
\bar{Q}^{}_{\Omega}H_{mn}
&=
\sqrt{2}ig[\varphi,\bar{\Lambda}_{mn}]
-\sqrt{2}g\Omega^{p}D_{p}\bar{\Lambda}_{mn}
+\sqrt{2}g(\Omega_{mp}\bar{\Lambda}_{pn}
-\Omega_{np}\bar{\Lambda}_{pm}). 
\label{defBRST1}
\end{align}
Here we have introduced an anti-self-dual auxiliary field $H_{mn}$ 
to make the supercharge $\bar{Q}_{\Omega}$ nilpotent off-shell 
up to the gauge transformation by $2\sqrt{2}\varphi$ 
and the $U(1)^{2}$ rotation by $2\sqrt{2}\Omega_{mn}$, 
which is the Cartan subgroup of the Lorentz group. 
We add the Gaussian term of $H_{mn}$ to the Lagrangian as 
\begin{gather}
\hat{\mathcal{L}}_{\Omega}=\mathcal{L}_{\Omega}+
\frac{1}{\kappa}\mathrm{Tr}\biggl[
-\frac{1}{2}(H_{mn}+F^{-}_{mn})^{2}
\biggr]. 
\label{omega_lag2}
\end{gather}
One can show that the Lagrangian \eqref{omega_lag2} 
is reduced to \eqref{omega_lag} 
by integrating out $H_{mn}$ and is invariant under the deformed
transformation \eqref{defBRST1}. 
The Lagrangian \eqref{omega_lag2} is written as 
the $\bar{Q}^{}_{\Omega}$-exact form up to the topological term
\begin{gather}
\hat{\mathcal{L}}_{\Omega}=\bar{Q}^{}_{\Omega}\Xi
+\biggl(\frac{8\pi^2}{g^2}-i\theta\biggr)
\frac{g^{2}}{32\pi^{2}}F_{mn}\tilde{F}^{mn}, 
\end{gather}
where $\Xi$ is given by 
\begin{align}
\Xi
&=
\frac{1}{\kappa}\textrm{Tr}\biggl[
-\frac{1}{2}F_{mn}^{-}\bar{\Lambda}^{mn}
-\frac{1}{4}H_{mn}\bar{\Lambda}^{mn}
-\frac{1}{2\sqrt{2}}\Lambda^{m}
(D_{m}\bar{\varphi}-gF_{mn}\bar{\Omega}^{n})
\notag\\
&\qquad\qquad\quad{}
+\frac{i}{4}g\bar{\Lambda}
\bigl([\varphi,\bar{\varphi}]
+i\Omega^{m}D_{m}\bar{\varphi}-i\bar{\Omega}^{m}D_{m}\varphi
+ig\bar{\Omega}^{m}\Omega^{n}F_{mn}
\bigr)\biggr]. 
\end{align}

The formula (\ref{defBRST1}) does not look like the one in \cite{NeOk}. 
However, \eqref{defBRST1} can be rewritten by introducing the operator 
$\Phi=\varphi+i\Omega^{m}D_{m}$ and 
$\bar{\Phi}=\bar{\varphi}+i\bar{\Omega}^{m}D_{m}$ as 
\begin{align}
\bar{Q}^{}_{\Omega}A_{m}
&=
\Lambda_{m},
\notag\\
\bar{Q}^{}_{\Omega}\Lambda_{m}
&=
-2\sqrt{2}(D_{m}\Phi+i\Omega_{mn}D_{n}),
\notag\\
\bar{Q}^{}_{\Omega}\bar{\Lambda}
&=
-2ig[\Phi,\bar{\Phi}],
\notag\\
\bar{Q}^{}_{\Omega}\bar{\Lambda}_{mn}
&=
2H_{mn},
\notag\\
\bar{Q}^{}_{\Omega}\Phi
&=
0,
\notag\\
\bar{Q}^{}_{\Omega}\bar{\Phi}
&=
-\sqrt{2}\bar{\Lambda},
\notag\\
\bar{Q}^{}_{\Omega}H_{mn}
&=
\sqrt{2}ig[\Phi,\bar{\Lambda}_{mn}]
+\sqrt{2}g(\Omega_{mp}\bar{\Lambda}_{pn}
-\Omega_{np}\bar{\Lambda}_{pm}).
\label{defBRST1a}
\end{align}
Then \eqref{defBRST1a} corresponds to the twisted supersymmetry transformation 
given in \cite{NeOk} though 
the spacetime noncommutativity is included in \cite{NeOk}. 
We note that in terms of $\Phi$ and $\bar{\Phi}$, the deformed supersymmetry 
transformation \eqref{defBRST1a} depends only on $\Omega_{mn}$, while 
$\Xi$ depends only on $\bar{\Omega}_{mn}$. 
This topological structure also appears in 
the instanton effective action because the deformed 
supersymmetry acts also on the instanton zero-modes.

\section{Instanton calculus in $\Omega$-background}
In this section, we study the instanton calculus for 
$\mathcal{N}=2$ super Yang-Mills theory in the $\Omega$-background. 
We will perform the ADHM construction of instantons in 
the $\Omega$-background and show that the instanton effective action 
is the same as that in \cite{Ne}, 
which is exact with respect to the deformed supercharge. 

We consider the coulomb branch of the theory where the scalar fields 
have vacuum expectation values (VEVs) 
$\langle\varphi\rangle=\phi^{0}$, 
$\langle\bar{\varphi}\rangle=\bar{\phi}^{0}$. 
Here $\phi^0$, $\bar{\phi}^0$ are diagonal matrices. 
In this case, we should consider 
the constrained instantons (see review \cite{DoHoKhMa}). 
We will expand the fields and the equations of motion in the coupling $g$, 
then solve the equations at the leading order in $g$. 
The expansion in $g$ is reliable when the VEVs 
$\phi^{0}$ and $\bar{\phi}^{0}$ are large. 

We consider the case of the self-dual instanton background. 
The anti-self-dual case can be studied similarly. 
In the self-dual instanton background, the fields are expanded in $g$ as 
\begin{align}
A_{m} 
&=
g^{-1} A^{(0)}_{m} + g A_{m}^{(1)} + \cdots, 
\notag\\
\Lambda^{I}
&=
g^{- \frac{1}{2}} \Lambda^{(0)I} + g^{\frac{3}{2}} \Lambda^{(1)I} + \cdots, 
\notag\\
\varphi 
&=
g^{0} \varphi^{(0)} + g^2 \varphi^{(1)} + \cdots, 
\notag\\
\bar{\varphi}
&=
g^0 \bar{\varphi}^{(0)} + g^2 \bar{\varphi}^{(1)} + \cdots, 
\notag\\
\bar{\Lambda}_{I}
&=
g^{\frac{1}{2}} \bar{\Lambda}^{(0)}_I
+ g^{\frac{5}{2}} \bar{\Lambda}^{(1)}_I + \cdots. 
\label{N2SD5}
\end{align}
The equations of motion at the leading order in $g$ are 
\begin{align}
&F_{mn}^{(0)-}=0,\quad \nabla^{m}F_{mn}^{(0)}=0,
\label{sdeq1}
\\
&\bar{\sigma}^{m\dot{\alpha}\alpha}\nabla_{m}\Lambda^{(0)I}_{\alpha}=0,
\label{sdeq4}
\\
&\nabla^{2}\varphi^{(0)}-\sqrt{2}i\Lambda^{(0)I}\Lambda^{(0)}_{I}
-\nabla^{m}(F^{(0)}_{mn}\Omega^{n})=0,
\label{scalar_eq}
\\
&\nabla^{2}\bar{\varphi}^{(0)}-\nabla^{m}(F^{(0)}_{mn}\bar{\Omega}^{n})=0,
\label{scalar_eq2}
\\
&\sigma^{m}_{\alpha\dot{\alpha}}\nabla_{m}\bar{\Lambda}^{(0)\dot{\alpha}}_{I}
-\sqrt{2}i[\bar{\varphi}^{(0)}, \Lambda^{(0)}_{\alpha I}]
\notag\\
&\qquad{}
+\sqrt{2}\bar{\Omega}^{m}\nabla_{m}\Lambda^{(0)}_{\alpha I}
-\frac{1}{\sqrt{2}}\bar{\Omega}^{+}_{mn}(\sigma^{mn})_{\alpha}{}^{\beta}
\Lambda^{(0)}_{\beta I}
-\sqrt{2}\bar{\mathcal{A}}^{J}{}_{I}\Lambda^{(0)}_{\alpha I}
=0,
\label{sdeq5}
\end{align}
where
$\nabla_{m}=\partial_{m}+i[A^{(0)}_{m},\ast]$ is
the covariant derivative in the instanton background. 
The deformation parameters 
$\Omega^{+mn}$ and $\bar{\Omega}^{+mn}$ are the self-dual part of 
$\Omega^{mn}$ and $\bar{\Omega}^{mn}$ respectively. 

The solution to \eqref{sdeq1}--\eqref{scalar_eq2} 
with the instanton number $k$ 
can be obtained by the ADHM construction \cite{AtHiDrMa,DoHoKhMa}. 
We introduce the $(N+2k)\times 2k$ matrix $\Delta_{\dot{\alpha}}$ defined by 
\begin{equation}
\Delta_{\dot{\alpha}}=a_{\dot{\alpha}}
+b^{\beta}\sigma_{m\beta\dot{\alpha}}x^{m}
=\binom{w_{\dot{\alpha}}}
{a'_{\alpha\dot{\alpha}}+\sigma_{m\alpha\dot{\alpha}}x^{m}}. 
\end{equation}
Here the parameters 
$a'_{m}=\frac{1}{2}\bar{\sigma}_{m}^{\dot{\alpha}\alpha}
a'_{\alpha\dot{\alpha}}$ and $w_{\dot{\alpha}}$ satisfy 
the bosonic ADHM constraints 
\begin{equation}
(\tau^{c})^{\dot{\alpha}}_{~\dot{\beta}}
(\bar{w}^{\dot{\beta}}w_{\dot{\alpha}}
+\bar{a}^{\prime\dot{\beta}\alpha}a'_{\alpha\dot{\alpha}})=0,\quad 
a'_{m}=\bar{a}'_{m}.
\label{ADHM}
\end{equation}
We also introduce the $(N+2k)\times N$ matrix $U$ which satisfies 
\begin{gather}
\bar{\Delta}^{\dot{\alpha}}U=0,\quad \bar{U}U=\boldsymbol{1}_{N},\quad 
U\bar{U}+\Delta_{\dot{\alpha}}f\bar{\Delta}^{\dot{\alpha}}=
\boldsymbol{1}_{N+2k}, 
\end{gather}
where $\boldsymbol{1}_{N}$ denotes the $N\times N$ identity matrix. 
The $k \times k$ matrix $f$ is expressed by $a'_{m}$ and $w_{\dot{\alpha}}$ as 
\begin{gather}
f=\biggl[\frac{1}{2}\bar{w}^{\dot{\alpha}}w_{\dot{\alpha}}
+(x_{m}+a'_{m})(x^{m}+a^{\prime m})\biggr]^{-1}.
\end{gather}
Then the self-dual gauge field $A^{(0)}_{m}$ is constructed as 
\begin{gather}
A^{(0)}_{m}=-i\bar{U}\partial_{m}U.  
\label{sol1}
\end{gather}
The Dirac equation in the instanton background \eqref{sdeq4} is solved as 
\begin{gather}
\Lambda^{(0)I}_{\alpha}=
\bar{U}(\mathcal{M}^{I}f\bar{b}_{\alpha}-b_{\alpha}f\bar{\mathcal{M}}^{I})U, 
\label{sol4}
\end{gather}
where 
$\mathcal{M}^{I}=(\mu^{I}\ \mathcal{M}^{\prime I}_{\alpha})^{\mathrm{T}}$ 
is the $(N+2k)\times k$ constant Grassmann-odd matrix 
which satisfies the fermionic ADHM constraints 
\begin{equation}
\bar{\mu}^{I}w_{\dot{\alpha}}+\bar{w}_{\dot{\alpha}}\mu^{I}
+[\mathcal{M}^{\prime\alpha I},a'_{\alpha\dot{\alpha}}]=0,
\quad \mathcal{M}^{\prime I}_{\alpha}=\bar{\mathcal{M}}^{\prime I}_{\alpha}.
\label{fADHM}
\end{equation}
The parameters 
$a'_{m}$, $w^{}_{\dot{\alpha}}$, 
$\mathcal{M}^{\prime I}_{\alpha}$ and $\mu^{I}$ are called 
the ADHM moduli. 
The equations \eqref{scalar_eq} and \eqref{scalar_eq2} for scalar fields 
are solved as 
\begin{align}
\varphi^{(0)}&=-i\frac{\sqrt{2}}{4}\epsilon_{IJ}
\bar{U}\mathcal{M}^{I}f\bar{\mathcal{M}}^{J}U
+\bar{U}\begin{pmatrix}
 \phi^{0} & 0 \\
 0 & \chi\boldsymbol{1}_{2}-i\boldsymbol{1}_{k}\Omega^{+}
 \end{pmatrix}U,
\label{sol2}
\\
\bar{\varphi}^{(0)}&=
\bar{U}\begin{pmatrix}
 \bar{\phi^{0}} & 0 \\
 0 & \bar{\chi}\boldsymbol{1}_{2}-i\boldsymbol{1}_{k}\bar{\Omega}^{+}
 \end{pmatrix}U,
\label{sol3}
\end{align}
where the $k\times k$ matrices $\chi$ and $\bar{\chi}$ satisfy 
\begin{align}
\boldsymbol{L}\chi
&=
i\frac{\sqrt{2}}{4}\epsilon_{IJ}
\bar{\mathcal{M}}^{I}\mathcal{M}^{J}
+\bar{w}^{\dot{\alpha}}\phi^{0} w_{\dot{\alpha}}
-i\Omega^{+mn}[a'_{m},a'_{n}],
\label{chi}
\\
\boldsymbol{L}\bar{\chi}
&=
\bar{w}^{\dot{\alpha}}\bar{\phi}^{0} w_{\dot{\alpha}}
-i\bar{\Omega}^{+mn}[a'_{m},a'_{n}]. 
\label{barchi}
\end{align}
Here $\boldsymbol{L}$ is defined by 
\begin{gather}
\boldsymbol{L}
=
\frac{1}{2}\bigl\{\bar{w}^{\dot{\alpha}}w_{\dot{\alpha}},\ast\bigr\}
+\bigl[a^{\prime m},[a'_{m},\ast]\bigr].
\end{gather}
$\Omega^{+}$ and $\bar{\Omega}^{+}$ are $2\times 2$ matrices 
whose components are given by 
\begin{gather}
(\Omega^{+})_{\alpha}{}^{\beta}
=\frac{1}{2}\Omega^{+}_{mn}(\sigma^{mn})_{\alpha}{}^{\beta},
\quad
(\bar{\Omega}^{+})_{\alpha}{}^{\beta}
=\frac{1}{2}\bar{\Omega}^{+}_{mn}(\sigma^{mn})_{\alpha}{}^{\beta}.
\end{gather}
We note that the ADHM constraints \eqref{ADHM} and 
\eqref{fADHM} are not deformed by 
the $\Omega$-background. The deformation appears only in 
the solution \eqref{sol2} and \eqref{sol3} of the scalar fields. 

By substituting the expansion of the fields \eqref{N2SD5} 
into the action $S=\int d^{4}x\,\mathcal{L}_{\Omega}$ 
and using the equations of motion at the leading order in $g$ 
\eqref{sdeq1}--\eqref{sdeq5}, 
the action is expanded as 
\begin{gather}
S=\biggl(\frac{8\pi^2}{g^2}-i\theta\biggr)k+g^{0}S^{(0)}
+\mathcal{O}(g^2), 
\label{classical_g-expansion}
\end{gather}
where $S^{(0)}$ can be written as 
\begin{align}
S^{(0)}
&=
\int d^{4}x\,
\frac{1}{\kappa}\textrm{Tr}\biggl[
(\nabla_{m}\varphi^{(0)}-F^{(0)}_{mn}\Omega^{n})
(\nabla^{m}\bar{\varphi}^{(0)}-F^{(0)mp}\bar{\Omega}_{p})
-\frac{i}{\sqrt{2}}\Lambda^{(0)I}[\bar{\varphi},\Lambda^{(0)}_{I}]
\notag\\
&\qquad\qquad{}
+\frac{1}{\sqrt{2}}\bar{\Omega}^{m}\Lambda^{(0)I}
\nabla_{m}\Lambda^{(0)}_{I}
-\frac{1}{2\sqrt{2}}\bar{\Omega}^{+}_{mn}\Lambda^{(0)I}
\sigma^{mn}\Lambda^{(0)}_{I}
-\frac{1}{\sqrt{2}}\bar{\mathcal{A}}^{J}{}_{I}
\Lambda^{(0)I}\Lambda^{(0)}_{J}
\biggr].
\label{insteff1}
\end{align} 
Note that $S^{(0)}$ does not depend on $\bar{\Lambda}_{I}^{(0)}$.  
Therefore we do not need to solve \eqref{sdeq5} at the leading order in $g$. 
We now evaluate the integral \eqref{insteff1} and 
express it in terms of the ADHM moduli. 
Using the formulae 
\begin{align}
\textrm{Tr}\bigl[
F^{(0)}_{mn}\Omega^{n}\nabla^{m}\bar{\varphi}^{(0)}
\bigr]
&=
\textrm{Tr}\bigl[
\nabla^{m}(F^{(0)}_{mn}\Omega^{n}\bar{\varphi}^{(0)})
+\Omega^{+mn}F^{(0)}_{mn}\bar{\varphi}^{(0)}
\bigr],
\\[2mm]
\textrm{Tr}\bigl[
F^{(0)}_{mn}\bar{\Omega}^{n}\nabla^{m}\varphi^{(0)}
\bigr]
&=
\textrm{Tr}\bigl[
\nabla^{m}(F^{(0)}_{mn}\bar{\Omega}^{n}\varphi^{(0)})
+\bar{\Omega}^{+mn}F^{(0)}_{mn}\varphi^{(0)}
\bigr],
\end{align}
\begin{align}
\textrm{Tr}\biggl[
\frac{1}{\sqrt{2}}\bar{\Omega}^{m}\Lambda^{(0)I}\nabla_{m}\Lambda^{(0)}_{I}
\biggr]
&=
\textrm{Tr}\biggl[
\frac{1}{\sqrt{2}}\nabla^{m}\Bigl(\bar{\Omega}^{n}\Lambda^{(0)I}
\sigma_{mn}\Lambda^{(0)}_{I}\Bigr)
+
\frac{1}{\sqrt{2}}\bar{\Omega}^{+}_{mn}\Lambda^{(0)I}
\sigma^{mn}\Lambda^{(0)}_{I}
\biggr],
\label{f1}
\end{align}
we can decompose $S^{(0)}$ into three parts as 
\begin{align}
S^{(0)}
&=
S_{\mathrm{I}}+S_{\mathrm{II}}+S_{\mathrm{III}}, \label{eq:s0}
\\
S_{\mathrm{I}}
&=
\int\! d^{4}x\,\frac{1}{\kappa}\textrm{Tr}\biggl[
\nabla_{m}\varphi^{(0)} \nabla^{m}\bar{\varphi}^{(0)}
-\Omega^{+mn}F^{(0)}_{mn}\bar{\varphi}^{(0)}
-\bar{\Omega}^{+mn}F^{(0)}_{mn}\varphi^{(0)}
\notag\\
&\qquad\qquad\qquad{}
-\frac{i}{\sqrt{2}}\Lambda^{(0)I}
[\bar{\varphi}^{(0)},\Lambda^{(0)}_{I}]
+\frac{1}{2\sqrt{2}}\bar{\Omega}^{+}_{mn}\Lambda^{(0)I}
\sigma^{mn}\Lambda^{(0)}_{I}
\biggr],
\\
S_{\mathrm{II}}
&=
\int\! d^{4}x\,\frac{1}{\kappa}\textrm{Tr}\biggl[
F^{(0)}_{mn}\Omega^{n}F^{(0)mp}\bar{\Omega}_{p}
-\frac{1}{\sqrt{2}}\bar{\mathcal{A}}^{J}{}_{I}
\Lambda^{(0)I}\Lambda^{(0)}_{J}
\biggr],
\\
S_{\mathrm{III}}
&=
\int\! d^{4}x\,\frac{1}{\kappa}\textrm{Tr}\Bigl[
-\nabla^{m}\bigl(F^{(0)}_{mn}\Omega^{n}\bar{\varphi}^{(0)}
+F^{(0)}_{mn}\bar{\Omega}^{n}\varphi^{(0)}\bigr)
\Bigr]. 
\end{align}
Note that the first term of the right hand side in \eqref{f1} does not 
contribute to the integral. We will evaluate 
$S_{\mathrm{I}}$, $S_{\mathrm{II}}$ and $S_{\mathrm{III}}$ separately. 
The first term 
$S_{\mathrm{I}}$ takes the same form which we evaluated in our previous paper 
\cite{ItNaSa2} since $S_{\mathrm{I}}$ depends only on the self-dual part of 
$\Omega_{mn}$ and $\bar{\Omega}_{mn}$. 
\begin{align}
S_{\mathrm{I}}
&=
\frac{4\pi^{2}}{\kappa}\mathrm{tr}_{k}\biggl[
\Bigl(
\bar{w}^{\dot{\alpha}}\bar{\phi}^{0} w_{\dot{\alpha}}
-i\bar{\Omega}^{+mn}[a'_{m},a'_{n}]
\Bigr)
\boldsymbol{L}^{-1}
\biggl(
i\frac{\sqrt{2}}{4}\epsilon_{IJ}
\bar{\mathcal{M}}^{I}\mathcal{M}^{J}
+\bar{w}^{\dot{\alpha}}\phi^{0} w_{\dot{\alpha}}
-i\Omega^{+mn}[a'_{m},a'_{n}]
\biggr)
\notag\\
&\qquad\quad{}
+i\frac{\sqrt{2}}{4}\epsilon_{IJ}\bar{\mu}^{I}\bar{\phi}^{0}\mu^{J}
+\frac{\sqrt{2}}{16}\bar{\Omega}^{+}_{mn}\epsilon_{IJ}
\mathcal{M}^{\prime \alpha I}(\sigma^{mn})_{\alpha}{}^{\beta}
\mathcal{M}^{\prime J}_{\beta}
-\frac{1}{4}\Omega^{+mn}\bar{\Omega}^{+}_{mn}
\bar{w}^{\dot{\alpha}}w_{\dot{\alpha}}
\biggr],
\label{s1a}
\end{align}
where $\mathrm{tr}_{k}$ denotes the trace of a $k\times k$ matrix. 
Here we expand the $k\times k$ matrix by the $U(k)$ generators 
$t^{U}$ $(U=1,2,\ldots, k^{2})$ normalized as 
$\mathrm{tr}_{k}[t^{U}t^{V}]=\kappa\delta^{UV}$ with the same constant
$\kappa$ as that for the $U(N)$ generators. 
The second term 
$S_{\mathrm{II}}$ can be evaluated using Osborn's formula \cite{Os} 
and Corrigan's inner product formula \cite{DoKhMa, DoHoKhMa} as
\begin{align}
S_{\mathrm{II}}
&=
\frac{2\pi^{2}}{\kappa}\mathrm{tr}_{k}\biggl[
\frac{1}{2}\Omega^{mn}\bar{\Omega}^{}_{mn}
\bar{w}^{\dot{\alpha}}w_{\dot{\alpha}}
+2\Omega^{mn}\bar{\Omega}_{mp}a'_{n}a^{\prime p}
+\frac{1}{\sqrt{2}}
\bar{\mathcal{A}}^{J}{}_{I}\biggl(\bar{\mu}^{I}\mu^{}_{J}
+\frac{1}{2}\mathcal{M}^{\prime \alpha I}\mathcal{M}'_{\alpha J}\biggr)
\biggr].
\end{align}
Since $S_{\mathrm{III}}$ is a total derivative, we can calculate it 
using the asymptotic behavior of $F^{(0)}_{mn}$ at large $|x|$: 
\begin{gather}
F^{(0)}_{mn}\sim -4i|x|^{-6}x^{p}x^{q}
w_{\dot{\alpha}}\bar{\sigma}_{p}^{\dot{\alpha}\alpha}
(\sigma_{mn})_{\alpha}{}^{\beta}
\sigma_{q\beta\dot{\beta}}\bar{w}^{\dot{\beta}}
\quad \mbox{for} \quad |x|\to\infty. 
\end{gather}
Then $S_{\mathrm{III}}$ is obtained as follows: 
\begin{align}
S_{\mathrm{III}}=\frac{2\pi^{2}}{\kappa}\mathrm{tr}_{k}\Bigl[
i\Omega^{-}_{mn}(\bar{\sigma}^{mn})^{\dot{\alpha}}{}_{\dot{\beta}}
\bar{w}^{\dot{\beta}}\bar{\phi}^{0}w_{\dot{\alpha}}
+i\bar{\Omega}^{-}_{mn}
(\bar{\sigma}^{mn})^{\dot{\alpha}}{}_{\dot{\beta}}
\bar{w}^{\dot{\beta}}\phi^{0} w_{\dot{\alpha}}
\Bigr].
\end{align}

In \eqref{eq:s0} the ADHM moduli obey the ADHM constraints. 
In the path integral formulation around the instanton solution, one can 
introduce new auxiliary 
variables 
such that the ADHM moduli become 
independent and the $\boldsymbol{L}^{-1}$ terms in \eqref{eq:s0} become 
Gaussian.
It turns out that $S^{(0)}$ 
is obtained from the new action
$S^{(0)}_{\mathrm{eff}}$ by integrating out the auxiliary 
variables, 
which is given by 
\begin{align}
S^{(0)}_{\mathrm{eff}}
&=
\frac{2\pi^{2}}{\kappa}\mathrm{tr}_{k}\biggl[
-2\Bigl([\bar{\chi},a'_{m}]-i\bar{\Omega}_{mn}a^{\prime n}\Bigr)
\Bigl([\chi,a^{\prime m}]-i\Omega^{mp}a'_{p}\Bigr)
\notag\\
&\qquad\qquad\quad{}
+(\bar{\chi}\bar{w}^{\dot{\alpha}}-\bar{w}^{\dot{\alpha}}\bar{\phi}^{0})
(w_{\dot{\alpha}}\chi-\phi^{0} w_{\dot{\alpha}})
+(\chi\bar{w}^{\dot{\alpha}}-\bar{w}^{\dot{\alpha}}\phi^{0})
(w_{\dot{\alpha}}\bar{\chi}-\bar{\phi}^{0}w_{\dot{\alpha}})
\notag\\
&\qquad\qquad\quad{}
+\frac{i}{2\sqrt{2}}\mathcal{M}^{\prime\alpha I}\epsilon_{IJ}
\biggl([\bar{\chi},\mathcal{M}_{\alpha}^{\prime J}]
-\frac{i}{2}\bar{\Omega}_{mn}^{+}(\sigma^{mn})_{\alpha}{}^{\beta}
\mathcal{M}^{\prime J}_{\beta}\biggr)
\notag\\
&\qquad\qquad\quad{}
-\frac{i}{\sqrt{2}}\bar{\mu}^{I}\epsilon_{IJ}(\mu^{J}\bar{\chi}
-\bar{\phi}^{0}\mu^{J})
+\frac{1}{\sqrt{2}}
\bar{\mathcal{A}}^{J}{}_{I}\biggl(\bar{\mu}^{I}\mu^{}_{J}
+\frac{1}{2}\mathcal{M}^{\prime \alpha I}\mathcal{M}'_{\alpha J}\biggr)
\notag\\
&\qquad\qquad\quad{}
+i\Omega^{-}_{mn}(\bar{\sigma}^{mn})^{\dot{\alpha}}{}_{\dot{\beta}}
\bar{w}^{\dot{\beta}}\bar{\phi}^{0}w_{\dot{\alpha}}
+i\bar{\Omega}^{-}_{mn}
(\bar{\sigma}^{mn})^{\dot{\alpha}}{}_{\dot{\beta}}
\bar{w}^{\dot{\beta}}\phi^{0} w_{\dot{\alpha}}
\notag\\
&\qquad\qquad\quad{}
-2i\Omega^{-mn}\bar{\chi}[a'_{m},a'_{n}]
-2i\bar{\Omega}^{-mn}\chi[a'_{m},a'_{n}]
+\frac{1}{2}\Omega^{-mn}\bar{\Omega}^{-}_{mn}
\bar{w}^{\dot{\alpha}}w_{\dot{\alpha}}
\notag\\
&\qquad\qquad\quad{}
-i\bar{\psi}^{\dot{\alpha}}_{I}
\bigl(\bar{\mu}^{I}w_{\dot{\alpha}}+\bar{w}_{\dot{\alpha}}\mu^{I}
+[\mathcal{M}^{\prime\alpha I},a'_{\alpha\dot{\alpha}}]\bigr)
\notag\\
&\qquad\qquad\quad{}
+iD^{c}(\tau^{c})^{\dot{\alpha}}{}_{\dot{\beta}}
\bigl(\bar{w}^{\dot{\beta}}w_{\dot{\alpha}}
+\bar{a}^{\prime\dot{\beta}\alpha}a'_{\alpha\dot{\alpha}}\bigr)
\biggr].
\label{insteff0}
\end{align}
The action $S^{(0)}_{\mathrm{eff}}$ is called the instanton effective action. 
The equations of motion for auxiliary variables 
$\chi$, $\bar{\chi}$, $D^{c}$ and $\bar{\psi}^{\dot{\alpha}}_{I}$ 
from \eqref{insteff0} give \eqref{chi}, \eqref{barchi}, 
the ADHM constraints \eqref{ADHM} and \eqref{fADHM} respectively. 
Substituting them into \eqref{insteff0}, we obtain $S^{(0)}$. 

We redefine $D^{c}$ as 
\begin{gather}
D^{c}\to D^{c}-\frac{i}{2}\bar{\eta}^{c}_{mn}
(\bar{\Omega}^{-mn}\chi+\Omega^{-mn}\bar{\chi}). 
\end{gather}
Here we have defined the 't Hooft $\eta$-symbols 
$\eta^{c}_{mn}$ and $\bar{\eta}^{c}_{mn}$ as 
$\sigma_{mn}=\frac{i}{2}\tau^{c}\eta^{c}_{mn}$ and 
$\bar{\sigma}_{mn}=\frac{i}{2}\tau^{c}\bar{\eta}^{c}_{mn}$ respectively. 
Then $S^{(0)}_{\mathrm{eff}}$ can be rewritten as 
\begin{align}
S^{(0)}_{\mathrm{eff}}
&=
\frac{2\pi^{2}}{\kappa}\mathrm{tr}_{k}\biggl[
-2\Bigl([\bar{\chi},a'_{m}]-i\bar{\Omega}_{mn}a^{\prime n}\Bigr)
\Bigl([\chi,a^{\prime m}]-i\Omega^{mp}a'_{p}\Bigr)
\notag\\
&\qquad\qquad\quad{}
+\frac{i}{2\sqrt{2}}\mathcal{M}^{\prime\alpha I}
\biggl([\bar{\chi},\mathcal{M}'_{\alpha I}]
-\frac{i}{2}\bar{\Omega}_{mn}^{+}(\sigma^{mn})_{\alpha}{}^{\beta}
\mathcal{M}'_{\beta I}
-i\bar{\mathcal{A}}^{J}{}_{I}
\mathcal{M}'_{\alpha J}
\biggr)
\notag\\
&\qquad\qquad\quad{}
+\biggl(
\bar{\chi}\bar{w}^{\dot{\alpha}}-\bar{w}^{\dot{\alpha}}\bar{\phi}^{0}
-\frac{i}{2}\bar{\Omega}^{-}_{mn}
(\bar{\sigma}^{mn})^{\dot{\alpha}}{}_{\dot{\beta}}\bar{w}^{\dot{\beta}}
\biggr)
\biggl(
w_{\dot{\alpha}}\chi-\phi^{0} w_{\dot{\alpha}}
-\frac{i}{2}\Omega^{-}_{mn}
(\bar{\sigma}^{mn})^{\dot{\gamma}}{}_{\dot{\alpha}}w_{\dot{\gamma}}
\biggr)
\notag\\
&\qquad\qquad\quad{}
+\biggl(
\chi\bar{w}^{\dot{\alpha}}-\bar{w}^{\dot{\alpha}}\phi^{0}
-\frac{i}{2}\Omega^{-}_{mn}
(\bar{\sigma}^{mn})^{\dot{\alpha}}{}_{\dot{\beta}}\bar{w}^{\dot{\beta}}
\biggr)
\biggl(
w_{\dot{\alpha}}\bar{\chi}-\bar{\phi}^{0}w_{\dot{\alpha}}
-\frac{i}{2}\bar{\Omega}^{-}_{mn}
(\bar{\sigma}^{mn})^{\dot{\gamma}}{}_{\dot{\alpha}}w_{\dot{\gamma}}
\biggr)
\notag\\
&\qquad\qquad\quad{}
-\frac{i}{\sqrt{2}}\bar{\mu}^{I}\Bigl(
\mu^{}_{I}\bar{\chi}-\bar{\phi}^{0}\mu^{}_{I}
+i\bar{\mathcal{A}}^{J}{}_{I}\mu^{}_{J}
\Bigr)
\notag\\
&\qquad\qquad\quad{}
-i\bar{\psi}^{\dot{\alpha}}_{I}
\bigl(\bar{\mu}^{I}w_{\dot{\alpha}}+\bar{w}_{\dot{\alpha}}\mu^{I}
+[\mathcal{M}^{\prime\alpha I},a'_{\alpha\dot{\alpha}}]\bigr)
\notag\\
&\qquad\qquad\quad{}
+iD^{c}(\tau^{c})^{\dot{\alpha}}{}_{\dot{\beta}}
\biggl(\bar{w}^{\dot{\beta}}w_{\dot{\alpha}}
+\bar{a}^{\prime\dot{\beta}\alpha}a'_{\alpha\dot{\alpha}}
-\frac{1}{2}(\tau^{c})^{\dot{\beta}}{}_{\dot{\alpha}}\zeta^{c}
\biggr)
\biggr]. 
\label{insteff}
\end{align}
Here we have introduced 
the Fayet-Iliopoulos parameter $\zeta^{c}$
(spacetime noncommutativity in $\mathcal{N}=2$ super Yang-Mills theory) 
which resolves the small instanton singularity \cite{NeSc}. 
We can choose $\zeta^{c}$ as $\zeta^{1}=\zeta^{2}=0$, $\zeta^{3}\neq 0$. 
The instanton effective action \eqref{insteff} reduces to the undeformed one 
when all the deformation parameters vanish. 
It becomes the action in \cite{BiFrFuLe, ItNaSa2} when we choose
the self-dual $\Omega$-background without 
the R-symmetry Wilson line gauge field such as 
\begin{gather}
\Omega^{-}_{mn}=\bar{\Omega}^{-}_{mn}=\bar{\mathcal{A}}^{I}{}_{J}=0. 
\end{gather}
In this case we have four deformed supersymmetries \cite{ItNaSa2}. 
But for general $\Omega$-background these supersymmetries 
are broken as we observed in the case of the spacetime action. 
Note that when \eqref{condition1} is satisfied, 
$S^{(0)}_{\mathrm{eff}}$ preserves one supersymmetry 
for general $\Omega_{mn}$ and $\bar{\Omega}_{mn}$ \cite{Ne, NeOk}. 
This unbroken supersymmetry plays a role of equivariant BRST symmetry. 
As we have done in the previous section, we write down the symmetry using 
the topological twist. We change the variables as 
\begin{gather}
\mathcal{M}'_{\alpha I}
=
\sigma^{m}_{\alpha I}\mathcal{M}'_{m},\quad
\bar{\psi}^{\dot{\alpha}}_{I}
=
-\frac{1}{2\sqrt{2}}\delta^{\dot{\alpha}}_{I}\bar{\eta}
+\frac{1}{2}(\bar{\sigma}^{mn})^{\dot{\alpha}}{}_{I}\bar{\psi}_{mn},\quad
D^{c}
=
\frac{i}{4}\bar{\eta}^{c}_{mn}D^{mn}.
\end{gather}
Then the twisted supercharge $\bar{Q}^{}_{\Omega}$ acts on 
the ADHM moduli and the auxiliary variables as 
\begin{align}
\bar{Q}^{}_{\Omega}a'_{m}&=\mathcal{M}'_{m},
&
\bar{Q}^{}_{\Omega}\mathcal{M}'_{m}&=
-2\sqrt{2}i[\chi, a'_{m}]-2\sqrt{2}\Omega_{m}{}^{n}a^{\prime}_{n},
\notag\\
\bar{Q}^{}_{\Omega}w_{\dot{\alpha}}&=\mu_{\dot{\alpha}},
&
\bar{Q}^{}_{\Omega}\mu_{\dot{\alpha}}&=
2\sqrt{2}i(w_{\dot{\alpha}}\chi-\phi^{0} w_{\dot{\alpha}})
+\sqrt{2}\Omega^{-}_{mn}
(\bar{\sigma}^{mn})^{\dot{\beta}}{}_{\dot{\alpha}}w_{\dot{\beta}},
\notag\\
\bar{Q}^{}_{\Omega}\bar{w}^{\dot{\alpha}}&=\bar{\mu}^{\dot{\alpha}},
&
\bar{Q}^{}_{\Omega}\bar{\mu}^{\dot{\alpha}}&=
-2\sqrt{2}i(\chi\bar{w}^{\dot{\alpha}}-\bar{w}^{\dot{\alpha}}\phi^{0})
-\sqrt{2}\Omega^{-}_{mn}
(\bar{\sigma}^{mn})^{\dot{\alpha}}{}_{\dot{\beta}}\bar{w}^{\dot{\beta}},
\notag\\
\bar{Q}^{}_{\Omega}\chi&=0, & & 
\notag\\
\bar{Q}^{}_{\Omega}\bar{\chi}&=\bar{\eta},
&
\bar{Q}^{}_{\Omega}\bar{\eta}&=-2\sqrt{2}i[\chi,\bar{\chi}],
\notag\\
\bar{Q}^{}_{\Omega}\bar{\psi}_{mn}&=D_{mn},
&
\bar{Q}^{}_{\Omega}D_{mn}&=
-2\sqrt{2}i[\chi,\bar{\psi}_{mn}]
-2\sqrt{2}(\Omega^{-}_{mp}\bar{\psi}_{pn}
-\Omega^{-}_{np}\bar{\psi}_{pm}).
\label{defBRST2}
\end{align}
We can show that the instanton effective action $S^{(0)}_{\mathrm{eff}}$ 
is invariant under the twisted supersymmetry transformation \eqref{defBRST2} 
which is the same as that given in \cite{Ne, NeOk} 
(see Appendix \ref{rewrite}). 
Note that $\bar{Q}^{}_{\Omega}$ is nilpotent up to the $U(k)$ transformation 
by $2\sqrt{2}\chi$, the $U(1)^{N}$ transformation by $2\sqrt{2}\phi^{0}$ 
and the $U(1)^{2}$ rotation by $2\sqrt{2}\Omega_{mn}$. 
The action $S^{(0)}_{\mathrm{eff}}$ can be written as 
the $\bar{Q}^{}_{\Omega}$-exact form 
$S^{(0)}_{\mathrm{eff}}=\bar{Q}^{}_{\Omega}\hat{\Xi}$, 
where $\hat{\Xi}$ is given by 
\begin{align}
\hat{\Xi}
&=
\frac{2\pi^{2}}{\kappa}\mathrm{tr}_{k}\biggl[
-\frac{i}{2\sqrt{2}}
\biggl(\bar{\chi}\bar{w}^{\dot{\alpha}}-\bar{w}^{\dot{\alpha}}\bar{\phi}^{0}
-\frac{i}{2}\bar{\Omega}^{-}_{mn}
(\bar{\sigma}^{mn})^{\dot{\alpha}}{}_{\dot{\beta}}\bar{w}^{\dot{\beta}}
\biggr)\mu_{\dot{\alpha}}
\notag\\
&\qquad\qquad\quad{}
+\frac{i}{2\sqrt{2}}\bar{\mu}^{\dot{\alpha}}
\biggr(w_{\dot{\alpha}}\bar{\chi}-\bar{\phi}^{0}w_{\dot{\alpha}}
-\frac{i}{2}\bar{\Omega}^{-}_{mn}
(\bar{\sigma}^{mn})^{\dot{\beta}}{}_{\dot{\alpha}}w_{\dot{\beta}}
\biggr)
\notag\\
&\qquad\qquad\quad{}
-\frac{i}{\sqrt{2}}\mathcal{M}'_{m}
\bigl([\bar{\chi},a^{\prime m}]-i\bar{\Omega}^{mn}a'_{n}\bigr)
\notag\\
&\qquad\qquad\quad{}
+\frac{i}{2}\bar{\psi}^{mn}\biggl\{
(\bar{\sigma}_{mn})^{\dot{\alpha}}{}_{\dot{\beta}}
\bar{w}^{\dot{\beta}}w_{\dot{\alpha}}-2[a'_{m},a'_{n}]
-\frac{i}{2}\zeta^{c}\bar{\eta}^{c}_{mn}
\biggr\}
\biggr].
\label{defGF2}
\end{align}
This is the same as that given in \cite{Ne}. 
Note that $\bar{\Omega}_{mn}$ appears only in $\hat{\Xi}$ 
but not in $\bar{Q}^{}_{\Omega}$, 
and hence the variation of $S^{(0)}_{\mathrm{eff}}$ with respect to 
$\bar{\Omega}_{mn}$ is $\bar{Q}^{}_{\Omega}$-exact. 
This leads to the fact that 
the instanton partition function 
does not depend on $\bar{\Omega}_{mn}$.

\section{D$(-1)$-brane effective action in R-R background}
In this section, 
we derive the instanton effective action of 
$\mathcal{N} = 2$ super Yang-Mills theory 
in the 
$\Omega$-background from superstring theory.

\subsection{Undeformed D$(-1)$-brane effective action}
We begin with realizing undeformed $\mathcal{N}=2$ $U(N)$ super
Yang-Mills theory and its instantons from D-branes in type IIB
superstring theory. 
We consider $N$ fractional D3-branes located at 
the fixed point of the orbifold $\mathbb{C} \times 
\mathbb{C}^2/\mathbb{Z}_2$ \cite{BiFrFuLe}.
Instantons with topological 
number $k$ 
correspond to $k$ fractional D$(-1)$-branes embedded in 
the D3-brane world-volume. 
The zero-modes of the open strings that have at least one end point on 
the D$(-1)$-branes 
are the moduli parameters and the auxiliary 
variables in the ADHM construction of instantons 
\cite{Douglas}. 
On the other hand, the zero-modes of the open strings 
that have both end points on the D3-branes correspond to 
the vector multiplet in $\mathcal{N} = 2$ supersymmetric theory. 
We 
calculate disk amplitudes which contain massless states of the open strings 
and take the zero-slope limit $\alpha' \to 0$. Here $\alpha'$ is the 
Regge slope parameter.
We obtain the low-energy effective action of the D3/D$(-1)$-brane system, which 
turns out to be the instanton effective action in $\mathcal{N} = 2$ super Yang-Mills theory.

We study the corrections to the D($-1$) effective action in closed string backgrounds
from string disk amplitudes that contain the open and closed string 
vertex operators. We use the NSR formalism to calculate the disk amplitudes. 
Conventions and notations for 
world-sheet variables are summarized in Appendix B. 
A disk is realized as the upper half-plane 
whose boundary is 
the real axis.
The open string vertex operators are inserted on the real axis  parameterized by $y$ 
while the closed string vertex operators are in the upper-half plane 
parameterized by $z, \bar{z}$. 
The closed string vertex operators consist of products of left and 
right moving fields. We employ the doubling trick where right moving fields are
located on the lower-half plane and left and right fields are identified 
on the boundary.  
Since the internal space is orbifolded by the $\mathbb{Z}_2$ action, 
the vertex operators must be invariant under 
the $\mathbb{Z}_2$ transformation. 
The $\mathbb{Z}_2$ transformations on the string world-sheet fields are 
found in Appendix B. 
We denote the vertex operator for a field 
$\Psi$, which corresponds to a massless state of the open strings, 
in the $q$-picture by $V^{(q)}_{\Psi}$.
The vertex operator associated with a closed string background 
$\mathcal{F}$ is denoted by $V^{(-1/2,-1/2)}_{\mathcal{F}}$, 
where we fix its picture number
to $(-1/2,-1/2)$.
The disk amplitude which contains $n_o$ open string vertex operators 
$V^{(q_i)}_{\Psi_i}(y_i)$ and $n_c$ closed string vertex operators 
$V^{(-1/2,-1/2)}_{\mathcal{F}_j} (z_j,\bar{z}_j)$ is given by 
\begin{equation}
\langle \! \langle V^{(q_1)}_{\Psi_1}\cdots V^{(-1/2,-1/2)}_{\mathcal{F}_1}
\cdots \rangle \! \rangle
= C_0 \int \frac{\prod_{i=1}^{n_o} dy_i \prod_{j=1}^{n_c}
d z_j d \bar{z}_j}{dV_{CKG}}
\langle V^{(q_1)}_{\Psi_1} (y_1) \cdots V^{(-1/2,-1/2)}_{\mathcal{F}_1} 
(z_1,\bar{z_1}) \cdots \rangle.
\label{eq:disk_gen}
\end{equation}
Here, $C_0$ is the disk normalization factor which is given by
\begin{equation}
 C_0 = \frac{1}{2\pi^2 \alpha^{\prime 2}} \frac{1}{\kappa g_0^2},
\end{equation}
where 
$g_0 = (2\pi)^{-3/2} g_s^{1/2} \alpha^{\prime -1}$ 
is the D$(-1)$-brane coupling constant \cite{Po} 
and $g_s$ is the string coupling constant.
The factor $dV_{CKG}$ is an $SL(2,\mathbb{R})$-invariant volume factor to fix
three positions $x_1$, $x_2$ and $x_3$ among $y_i$, $z_j$,and 
$\bar{z}_j$'s, which is given by
\begin{equation}
 d V_{CKG}={d x_1 d x_2 d x_3\over
(x_1-x_2) (x_2-x_3) (x_3-x_1)}.
\end{equation}
We fix one position of an open string vertex operator 
to $y_1$ and two positions of a closed string vertex operator 
to $z, \bar{z}$.
Note that in the disk amplitude (\ref{eq:disk_gen}) the sum of the 
picture numbers
must be $-2$. 

The $\mathbb{Z}_2$-invariant open string vertex operators corresponding to 
the ADHM moduli, the auxiliary variables and the VEVs of the scalar 
fields are summarized in Table \ref{N2ADHM}, 
where we omit the normal ordering symbol throughout this paper.
\begin{table}[t]
\begin{center}
\begin{tabular}{|l||l|l|}
\hline
Brane sectors & Vertex Operators & Zero-modes \\
\hline \hline
D$(-1)$/D$(-1)$ &   $ \displaystyle V^{(-1)}_{a'} (y) = \pi (2 \pi 
 \alpha')^{\frac{1}{2}} g_0 \frac{a'_{m}}{\sqrt{2}} \psi^{m}  e^{- \phi} 
 (y) $ & ADHM moduli \\
            &   $ \displaystyle V_{\mathcal{M}}^{(- 1/2)} (y)= \pi (2 \pi \alpha')^{\frac{3}{4}} g_0
 \mathcal{M}^{\prime \alpha I}  S_{\alpha} S_I  e^{ - \frac{1}{2}  \phi} (y)$ &  \\
\cline{2-3}
            &   $ \displaystyle V_{\chi}^{(-1)} (y)= (2 \pi \alpha')^{\frac{1}{2}} 
\frac{\chi}{\sqrt{2}} \bar{\psi}  e^{- \phi } (y)$ &   Auxiliary variables\\
            &   $ \displaystyle V_{\bar{\chi}}^{(-1)} (y)= (2 \pi \alpha')^{\frac{1}{2}} 
\frac{\bar{\chi}}{\sqrt{2}} \psi  e^{- \phi } (y)$ &   \\
            &   $ \displaystyle V_{\bar{\psi}}^{(- 1/2)} (y)= 2 (2 \pi \alpha')^{\frac{3}{4}} 
 \bar{\psi}_{\dot{\alpha} I} S^{\dot{\alpha}}  S^I  e^{- \frac{1}{2} 
 \phi}  (y) $  & \\
            &   $ \displaystyle V^{(0)}_D (y) = 
2 (2 \pi \alpha') D_c \bar{\eta}^c_{mn} \psi^{n} \psi^{m} (y) $ &   \\
\hline
D3/D$(-1)$ &   $\displaystyle V_{w}^{(-1)} (y)= \pi (2 \pi \alpha')^{\frac{1}{2}} g_0 \frac{w_{\dot{\alpha}}}{2} 
\Delta S^{\dot{\alpha}} e^{- \phi } (y) $ & ADHM moduli \\
         &   $\displaystyle V^{(-1)}_{\bar{w}} (y) = \pi (2 \pi 
 \alpha')^{\frac{1}{2}} g_0 \frac{\bar{w}_{\dot{\alpha}}}{2} \overline{\Delta} 
 S^{\dot{\alpha}} e^{- \phi} (y)  $ &  \\
        &   $\displaystyle V_{\mu}^{(-1/2)} (y)= \pi (2 \pi \alpha')^{\frac{3}{4}} g_0 
\mu^I \Delta S_I e^{- \frac{1}{2} \phi } (y) $ &  \\
        &   $\displaystyle V_{\bar{\mu}}^{(-1/2)} (y)= \pi (2 \pi \alpha')^{\frac{3}{4}} g_0 
\bar{\mu}^I \overline{\Delta} S_I e^{- \frac{1}{2} \phi } (y) $ &  \\
\hline
D3/D3& $ \displaystyle \frac{}{} V^{(-1)}_{\phi} (y) = 
(2 \pi \alpha')^{\frac{1}{2}} \frac{\phi^0}{\sqrt{2}} \bar{\psi} e^{- 
\phi} (y)$ & VEVs \\
    &  $ \displaystyle \frac{}{} V^{(-1)}_{\bar{\phi}} (y) = 
(2 \pi \alpha')^{\frac{1}{2}} \frac{\bar{\phi}^0}{\sqrt{2}} \psi e^{- 
\phi} (y)$ & \\
\hline
\end{tabular}
\caption{Vertex operators for the ADHM moduli, the auxiliary variables 
and the VEVs.}
\label{N2ADHM}
\end{center}
\end{table}
Here $\psi^m, \psi, \bar{\psi}$ are world-sheet fermions, 
$\phi$ 
is the free boson for the bosonization of the bosonic ghost \cite{FrMaSh}, 
whose momentum in a vertex operator specifies the picture number. 
The fields $\Delta$ and $\bar{\Delta}$ are 
the twist fields which interchange the D3 and D$(-1)$ 
boundaries 
\cite{Twist}. 
Parallel D3-branes break the $SO(10)$ Lorentz symmetry down to $SO(4) \times 
SO(6)$. 
The ten-dimensional spin field is decomposed into the four-dimensional part
$S^{\alpha}, S_{\dot{\alpha}}$ and the six-dimensional part $S^A, S_A$ 
where $A=1, \cdots, 4$ is an $SO(6)$ spinor ($SU(4)$ vector) index. 
Once the internal space is orbifolded, the $SU(4)$ symmetry is 
broken to $SU(2) \times SU(2)$ and the $SU(4)$ index $A$ is 
decomposed into the $SU(2) \times SU(2)$ indices $I=1,2$ and $I'=3,4$.
The first $SU(2)$ corresponds to the R-symmetry $SU(2)_I$ in $\mathcal{N} = 2$ 
supersymmetry on the D3-branes while the second $SU(2)$ is 
associated with the other $\mathcal{N}=2$ supersymmetry which is broken on the D3-branes.

Note that in order to reproduce the 
undeformed instanton effective action in the zero-slope limit, 
some of the moduli should be rescaled by $g_0$ \cite{BiFrPeFuLeLi}. 
The powers of $\alpha'$ in the vertex operators are determined 
such that the zero-modes have the canonical dimensions.
The zero-modes of the D$(-1)$/D$(-1)$ strings belong to the adjoint representation of
$U(k)$ while the zero-modes of the D3/D$(-1)$ strings belong to the bi-fundamental representation 
of $U(k) \times U(N)$ gauge group. 
We use the same normalization of the $U(k)$ generators as in section 3.

In addition to the open string zero-modes in Table \ref{N2ADHM}, 
we introduce auxiliary fields 
$Y_m, Y^{\dagger}_m, X_{\dot{\alpha}}, \bar{X}_{\dot{\alpha}}, 
X_{\dot{\alpha}}^{\dagger}, \bar{X}_{\dot{\alpha}}^{\dagger}$ 
to disentangle higher point 
interactions in the instanton effective action \cite{BiFrFuLe}.
The vertex operators associated with these auxiliary fields 
are given in Table \ref{N2auxiliary}.
\begin{table}[t]
\begin{center}
\begin{tabular}{|l||l|l|}
\hline
Brane sectors &  Vertex Operators & Fields \\
\hline \hline
D$(-1)$/D$(-1)$ &   $ \displaystyle V_Y^{(0)} (y) = 4\pi (2 \pi \alpha') 
g_0 Y_{m} \psi^{m} \bar{\psi} (y) $ & Auxiliary fields \\
            &   $ \displaystyle V_{Y^{\dagger}}^{(0)} (y) = 4\pi (2 \pi \alpha') 
g_0 Y^{\dagger}_{m} \psi^{m} \psi (y) $ &  \\
\cline{1-2}
D3/D$(-1)$ &   $ \displaystyle V_X^{(0)} (y)= 2\sqrt{2} \pi 
(2 \pi \alpha') g_0 X_{\dot{\alpha}} \Delta S^{\dot{\alpha}} \bar{\psi} (y)$ & 
 \\
         &   $ \displaystyle V_{X^{\dagger}}^{(0)} (y)= 2\sqrt{2} \pi 
(2 \pi \alpha') g_0 X^{\dagger}_{\dot{\alpha}} \Delta S^{\dot{\alpha}} \psi (y)$ &  \\
            &  $ \displaystyle V_{\overline{X}}^{(0)} (y)= 
	    2\sqrt{2} \pi (2 \pi \alpha') g_0 \overline{X}_{\dot{\alpha}} 
\overline{\Delta} S^{\dot{\alpha}} \bar{\psi} (y) $ &   \\
            &  $ \displaystyle V_{\overline{X}^{\dagger}}^{(0)} (y)= 
	    2\sqrt{2} \pi (2 \pi \alpha') g_0 
	    \overline{X}^{\dagger}_{\dot{\alpha}} 
\overline{\Delta} S^{\dot{\alpha}} \psi (y) $ &  \\
\hline
\end{tabular}
\caption{Vertex operators for the auxiliary fields}
\label{N2auxiliary}
\end{center}
\end{table}

Amplitudes in the zero-slope limit
including the fields in Table \ref{N2ADHM} and \ref{N2auxiliary} have been 
calculated in \cite{BiFrFuLe}. 
The action $\tilde{S}_{\mathrm{str}}$ which reproduces these amplitudes is 
given by
\begin{eqnarray}
\tilde{S}_{\mathrm{str}} = \tilde{S} + S_{\mathrm{ADHM}},
\end{eqnarray}
where 
\begin{eqnarray}
\tilde{S} &=& 
\frac{2\pi^2}{\kappa} \mathrm{tr}_k
\bigg[
2 Y^m Y^{\dagger}_m - X_{\dot{\alpha}} \overline{X}^{\dagger \dot{\alpha}} 
- X^{\dagger}_{\dot{\alpha}} \overline{X}^{\dot{\alpha}} 
+ 2 Y^m [\bar{\chi}, a'_m] + 2 Y^{\dagger m} [\chi, a'_m]
\nonumber \\
& & \qquad \qquad 
- X_{\dot{\alpha}} (\bar{\chi} \bar{w}^{\dot{\alpha}} - 
\bar{w}^{\dot{\alpha}} \bar{\phi}^0) - X^{\dagger}_{\dot{\alpha}} (\chi 
\bar{w}^{\dot{\alpha}} - \bar{w}^{\dot{\alpha}} \phi^0) 
\nonumber \\
& & \qquad \qquad 
- (w_{\dot{\alpha}} \bar{\chi} - \bar{\phi}^0 w_{\dot{\alpha}}) \overline{X}^{\dot{\alpha}}
- (w_{\dot{\alpha}} \chi - \phi^0 w_{\dot{\alpha}}) 
\overline{X}^{\dagger \dot{\alpha}} 
\nonumber \\
& & \qquad \qquad 
+ \frac{\sqrt{2}}{2} i \epsilon_{IJ} \bar{\mu}^I (- \mu^J \bar{\chi} + 
\bar{\phi}^0 \mu^J) + \frac{\sqrt{2}}{4} i \epsilon_{IJ} 
\mathcal{M}^{\prime \alpha I} [\bar{\chi}, \mathcal{M}'_{\alpha} {}^J]
\bigg],
\end{eqnarray}
and $S_{\mathrm{ADHM}}$ corresponds to the Lagrange multiplier terms 
for the ADHM constraints:
\begin{eqnarray}
S_{\mathrm{ADHM}} &=& \frac{2\pi^2}{\kappa} 
\mathrm{tr}_k 
\big[
- i \bar{\psi}^{\dot{\alpha}}_I 
(\bar{\mu}^I w_{\dot{\alpha}} + \bar{w}_{\dot{\alpha}} \mu^I + 
[\mathcal{M}^{\prime \alpha I}, a'_{\alpha \dot{\alpha}}]) 
\nonumber \\
& & \qquad \qquad 
+ i D^c (\tau^c)^{\dot{\alpha}} {}_{\dot{\beta}} (\bar{w}^{\dot{\beta}} 
w_{\dot{\alpha}} + \bar{a}^{\prime \dot{\beta} \alpha} a'_{\alpha \dot{\alpha}})
\big].
\end{eqnarray}
Once we integrate out the auxiliary fields 
we recover the instanton effective action 
of $\mathcal{N} = 2$ super Yang-Mills theory in flat 
spacetime \cite{DoHoKhMa}.

\subsection{R-R 3-form background}
We now introduce a closed string background.
In \cite{BiFrFuLe},  
the case of the constant self-dual R-R 3-form 
background $\mathcal{F}_{mna} \ (m,n=1,\cdots, 4, a = 5,6)$
in type IIB string theory has been 
studied. 
By identifying the self-dual R-R 
backgrounds with the self-dual $\Omega$-background parameters
with $\Omega_{mn} = \Omega_{mn}^{+}, 
\bar{\Omega}_{mn} = \bar{\Omega}_{mn}^{+}$,
the authors showed that the deformed low-energy effective action 
of D3/D$(-1)$-branes coincides 
with the 
instanton effective action in $\mathcal{N} = 2$ super Yang-Mills theory 
in the self-dual $\Omega$-background.
In order to realize the deformed instanton effective action in the general 
$\Omega$-background, we need to consider more general R-R backgrounds. 

Before evaluating background corrections to 
the effective action of the D3/D$(-1)$-brane system, 
we discuss a classification of the type IIB R-R backgrounds in the 
presence of D3-branes. 
Ten-dimensional R-R field strengths are expressed by the bi-spinor form 
$\mathcal{F}^{\hat{\mathcal{A}} \hat{\mathcal{B}}}$ where $\hat{\mathcal{A}}, 
\hat{\mathcal{B}}$ are 16 component ten-dimensional spinor indices. 
Since the D3-branes break 
the $SO(10)$ Lorentz symmetry down to $SO(4) \times SO(6)$, 
the ten-dimensional R-R backgrounds are 
decomposed into the four sectors
\begin{eqnarray}
\mathcal{F}^{\hat{\mathcal{A}} \hat{\mathcal{B}}} = 
(\mathcal{F}^{\alpha \beta AB}, \mathcal{F}^{\alpha} {}_{\dot{\beta}} 
{}^A {}_B, \mathcal{F}_{\dot{\alpha}} {}^{\beta} {}_A {}^B, 
\mathcal{F}_{\dot{\alpha} \dot{\beta} AB} ).
\label{eq:R-R_decomposition}
\end{eqnarray}
Each part 
in the right hand side of (\ref{eq:R-R_decomposition}) 
contains 
the R-R 1,3 and 5-form field strengths and is 
rewritten as 
\begin{eqnarray}
& & \mathcal{F}^{\alpha \beta AB} 
= \epsilon^{\alpha \beta} (\Sigma^a)^{AB} \mathcal{F}_a 
+ \epsilon^{\alpha \beta} (\Sigma^a \bar{\Sigma}^b \Sigma^c)^{AB} \mathcal{F}_{abc} 
\nonumber \\ 
& & \qquad \qquad 
+ \epsilon^{\alpha \gamma} (\sigma^{mn})_{\gamma} {}^{\beta} (\Sigma^a)^{AB} \mathcal{F}_{mna}
+ \epsilon^{\alpha \gamma} (\sigma^{mn})_{\gamma} {}^{\beta} (\Sigma^a \bar{\Sigma}^b \Sigma^c)^{AB} \mathcal{F}_{mnabc}, 
\nonumber \\
& & \mathcal{F}^{\alpha} {}_{\dot{\beta}} {}^A {}_B 
= \epsilon^{\alpha \gamma} (\sigma^m)_{\gamma \dot{\beta}} 
\left(\delta^A {}_B \mathcal{F}_m + 
(\Sigma^a \bar{\Sigma}^b)^A {}_B \mathcal{F}_{mab} \right), 
\nonumber \\
& & \mathcal{F}_{\dot{\alpha}} {}^{\beta} {}_A {}^B 
= \epsilon_{\dot{\alpha} \dot{\gamma}} (\bar{\sigma}^m)^{\dot{\gamma} \beta} 
\left(
\delta_A {}^B \mathcal{F}_m + (\bar{\Sigma}^a \Sigma^b)_A {}^B 
\mathcal{F}_{mab} 
\right), 
\nonumber \\
& & \mathcal{F}_{\dot{\alpha} \dot{\beta} AB} 
= \epsilon_{\dot{\alpha} \dot{\beta}} (\bar{\Sigma}^a)_{AB} 
\mathcal{F}_a 
+ \epsilon_{\dot{\alpha} \dot{\beta}} 
(\bar{\Sigma}^a \bar{\Sigma}^b \Sigma^c)_{AB} \mathcal{F}_{abc} 
\nonumber \\
& & \qquad \qquad 
+ 
\epsilon_{\dot{\alpha} \dot{\gamma}} (\bar{\sigma}^{mn})^{\dot{\gamma}} {}_{\dot{\beta}} 
(\bar{\Sigma}^a)_{AB} 
\mathcal{F}_{mna} + 
\epsilon_{\dot{\alpha} \dot{\gamma}} 
(\bar{\sigma}^{mn})^{\dot{\gamma}} {}_{\dot{\beta}} 
(\bar{\Sigma}^a \Sigma^b \bar{\Sigma}^c)_{AB} \mathcal{F}_{mnabc},
\end{eqnarray}
where we have used the ten-dimensional self-duality of the R-R 5-form 
field strength.
Here, $m,n,p,q = 1,\cdots 4$, denote D3-brane world-volume directions while $a, 
b, c = 5, \cdots 10$, are transverse directions to the D3-branes. 
The six-dimensional sigma matrices are defined by 
$(\Sigma^a)^{AB} = (\eta^3, - i \bar{\eta}^3, \eta^2, - i \bar{\eta}^2, 
\eta^1, i \bar{\eta}^1)$, $(\bar{\Sigma}^a)_{AB}
= (- \eta^3, - i \bar{\eta}^3, - \eta^2, - i \bar{\eta}^2, - \eta^1, i \bar{\eta}^1)
$.

The deformation parameters 
$\Omega_{mn}$ and $\bar{\Omega}_{mn}$
are anti-symmetric 
and the R-symmetry Wilson lines
${\cal A}$, $\bar{\cal A}$ 
have no spacetime 
indices.
So we focus on constant R-R 3-form backgrounds with index structures 
$\mathcal{F}_{mna}$ and $\mathcal{F}_{abc}$. 
We note that the other 
R-R backgrounds, namely R-R 
1 and 5-forms, induce non(anti)commutativity in superspace 
\cite{BoGrNi, BeSe, BiFrPeLe, ItSa, ItKoSa} 
that provides other 
types of deformations of super Yang-Mills theories. 

In the bi-spinor notation, 
the R-R 3-form $\mathcal{F}_{mna}$ corresponds to
\begin{eqnarray}
\mathcal{F}^{(\alpha \beta)[AB]}, \qquad 
\mathcal{F}_{(\dot{\alpha} \dot{\beta})[AB]},
\label{eq:SA_background}
\end{eqnarray}
where the round parentheses $(\cdot \cdot)$ 
denote symmetrization of indices while 
the square bracket $[\cdot \cdot]$ stands for anti-symmetrization of indices. 
The first background in 
(\ref{eq:SA_background}) 
corresponds to the self-dual while the second one corresponds to 
the anti-self-dual part of the field strength
in four dimensions.
We call these (S,A)-type 
backgrounds.
On the other hand, $\mathcal{F}_{abc}$ corresponds to the following structures 
\begin{eqnarray}
\mathcal{F}^{[\alpha \beta](AB)}, \qquad \mathcal{F}_{[\dot{\alpha} \dot{\beta}](AB)}.
\end{eqnarray}
We call these (A,S)-type backgrounds.

Now we consider the case that the internal space is orbifolded as 
$\mathbb{C} \times \mathbb{C}^2/\mathbb{Z}_2$. 
The vertex operators for the R-R 3-form backgrounds 
that are invariant under the $\mathbb{Z}_2$ orbifold projection 
 are summarized in Table \ref{N2RRSA}, \ref{N2RRAS}.
Here we have used the doubling trick and replaced
the anti-holomorphic part of the vertex operator by the holomorphic one
with argument $\bar{z}$. 
\begin{table}[t]
\begin{center}
\begin{tabular}{|l||l|}
\hline
Backgrounds & Vertex operators \\
\hline \hline
      & 
$ \displaystyle \frac{}{} V^{(-1/2,-1/2)}_{\mathcal{F}(+)} (z,\bar{z}) = 
 (2 \pi \alpha') \mathcal{F}^{(\alpha \beta) [IJ]} S_{\alpha} S_I 
e^{-\frac{1}{2} \phi} (z) S_{\beta} S_J e^{- \frac{1}{2} \phi} (\bar{z}) $ \\
\cline{2-2}
(S,A)-type & 
$ \displaystyle \frac{}{} V^{(-1/2,-1/2)}_{\overline{\mathcal{F}} (+)} (z,\bar{z}) = 
 (2 \pi \alpha') \overline{\mathcal{F}}^{(\alpha \beta) [I'J']} S_{\alpha} S_{I'} 
e^{-\frac{1}{2} \phi} (z) S_{\beta} S_{J'} e^{- \frac{1}{2} \phi} (\bar{z}) $ \\
\cline{2-2}
      & 
$ \displaystyle \frac{}{} V^{(-1/2,-1/2)}_{\overline{\mathcal{F}} (-)} 
(z, \bar{z}) = 
(2 \pi \alpha') \overline{\mathcal{F}}_{(\dot{\alpha} \dot{\beta}) [IJ]} S^{\dot{\alpha}} S^I 
e^{-\frac{1}{2} \phi} (z) S^{\dot{\beta}} S^J e^{- \frac{1}{2} \phi} (\bar{z}) $ \\
\cline{2-2}
      & 
$ \displaystyle \frac{}{} V^{(-1/2,-1/2)}_{\mathcal{F} (-)} (z, \bar{z}) = 
(2 \pi \alpha') \mathcal{F}_{(\dot{\alpha} \dot{\beta}) [I'J']} S^{\dot{\alpha}} S^{I'} 
e^{-\frac{1}{2} \phi} (z) S^{\dot{\beta}} S^{J'} e^{- \frac{1}{2} \phi} (\bar{z}) $ \\
\hline
\end{tabular}
\caption{Vertex operators for the (S,A)-type backgrounds.}
\label{N2RRSA}
\end{center}
\end{table}
\begin{table}[t]
\begin{center}
\begin{tabular}{|l||l|l|}
\hline
Backgrounds & Vertex operators  \\
\hline \hline
      & 
$ \displaystyle \frac{}{} V^{(-1/2,-1/2)}_{\mathcal{F}} (z, \bar{z}) = 
(2 \pi \alpha') \mathcal{F}^{[\alpha \beta] (IJ)} S_{\alpha} S_I 
e^{-\frac{1}{2} \phi} (z) S_{\beta} S_J e^{- \frac{1}{2} \phi} (\bar{z}) $ \\
\cline{2-2}
(A,S)-type & 
$ \displaystyle \frac{}{} V^{(-1/2,-1/2)}_{\mathcal{F}'} (z, \bar{z}) = 
(2 \pi \alpha') \mathcal{F}^{\prime [\alpha \beta] (I'J')} S_{\alpha} S_{I'} 
e^{-\frac{1}{2} \phi} (z) S_{\beta} S_{J'} e^{- \frac{1}{2} \phi} (\bar{z}) $ \\
\cline{2-2}
      & 
$ \displaystyle \frac{}{} V^{(-1/2,-1/2)}_{\overline{\mathcal{F}}} (z, \bar{z}) = 
(2 \pi \alpha') \overline{\mathcal{F}}_{[\dot{\alpha} \dot{\beta}] (IJ)} S^{\dot{\alpha}} S^{I} 
e^{-\frac{1}{2} \phi} (z) S^{\dot{\beta}} S^{J} e^{- \frac{1}{2} \phi} (\bar{z}) $ \\
\cline{2-2}
      & 
$ \displaystyle \frac{}{} V^{(-1/2,-1/2)}_{\overline{\mathcal{F}}'} (z, \bar{z}) = 
(2 \pi \alpha') \overline{\mathcal{F}'}_{[\dot{\alpha} \dot{\beta}] (I'J')} S^{\dot{\alpha}} S^{I'} 
e^{-\frac{1}{2} \phi} (z) S^{\dot{\beta}} S^{J'} e^{- \frac{1}{2} \phi} (\bar{z}) $ \\
\hline
\end{tabular}
\caption{Vertex operators for the (A,S)-type backgrounds.}
\label{N2RRAS}
\end{center}
\end{table}

We look for amplitudes that are non-vanishing in the zero-slope limit 
and contain at least one background vertex operator. 
As we discussed in 
\cite{ItNaSa2, ItNaSaSa}, 
we consider the zero-slope limit 
with fixed $(2\pi \alpha')^{\frac{1}{2}} \mathcal{F}$. 
Here $\mathcal{F}$ is the (S,A) or (A,S)-type background.
In addition to the closed string vertex operator, some open string vertex operators are 
 attached on disks.
In order to cancel the overall factor $1/g_0^2$ 
of disk amplitudes in the zero-slope limit 
the following combinations of fields need to be inserted on disks:
\begin{eqnarray}
\mu \bar{\mu},
\ w \bar{w},
\ Y a', ,
\ Y^{\dagger} a',
\ \mathcal{M}' \mathcal{M}',
\ a' a', 
\ w \bar{X}, 
\ w \bar{X}^{\dagger},
\ \bar{w} X,
\ \bar{w} X^{\dagger}.
\label{comb}
\end{eqnarray}
Note that the limit $\alpha' \to 0$ corresponds to $g_0 \to \infty$ 
because we keep the D3-brane (Yang-Mills) coupling constant 
$g = (2\pi)^{\frac{1}{2}} g^{\frac{1}{2}}_s$ \cite{Po} finite.

First we examine disk amplitudes with three or higher open string vertex
operators and one closed string vertex operator. 
Note that if a vertex operator contains the twist field, 
there should appear as  a pair of $\Delta$ and $\bar{\Delta}$ 
in the amplitude.
Among these combinations, 
$w \bar{w}$ can be accompanied with $a'$ by the dimensional analysis. 
However, the amplitude that contains $w \bar{w} a' \mathcal{F}$ vanishes 
because the sum of the charges associated with the spin operators of 
the internal space is not zero.
The combination $a'a'$ can be accompanied with $\chi$ or $\bar{\chi}$ by the 
dimensional analysis. 
However, amplitudes involving $a' a' \chi \mathcal{F}$ and $a' a' 
\bar{\chi} \mathcal{F}$ are zero for the 
anti-self-dual (S,A)- and the (A,S)-type backgrounds.
On the other hand, these amplitudes are reducible for the self-dual (S,A)-type 
backgrounds, which give no new interaction terms in the effective action.
By a similar analysis for higher point amplitudes we find that 
we need to consider only three point amplitudes of the ADHM moduli, 
the auxiliary variables, the auxiliary fields and the backgrounds $\mathcal{F}$. 
In the following subsections, we calculate 
the amplitudes that contain the combinations (\ref{comb}) and 
one closed string vertex operator for each background separately. 

\subsection{Amplitudes with the (S,A) and (A,S)-type backgrounds}
In this subsection, we evaluate the disk amplitudes that contain 
 the (S,A) and (A,S)-type backgrounds.
Let us first consider the anti-self-dual (S,A)-type backgrounds.
We find that the non-zero amplitudes are given by
\begin{eqnarray}
& & 
\langle \! \langle 
V^{(0)}_Y V^{(-1)}_{a'} V^{(-1/2,-1/2)}_{\bar{\mathcal{F}}(-)} 
\rangle \! \rangle, \quad 
\langle \! \langle 
V^{(0)}_{Y^{\dagger}} V^{(-1)}_{a'} V^{(-1/2,-1/2)}_{\mathcal{F}(-)} 
\rangle \! \rangle, \quad 
\langle \! \langle 
V^{(0)}_X V^{(-1)}_{\bar{w}} V^{(-1/2,-1/2)}_{\bar{\mathcal{F}}(-)} 
\rangle \! \rangle, \nonumber \\
& &  
\langle \! \langle 
V^{(0)}_{X^{\dagger}} V^{(-1)}_{\bar{w}} V^{(-1/2,-1/2)}_{\mathcal{F}(-)} 
\rangle \! \rangle, \quad 
\langle \! \langle 
V^{(0)}_{\bar{X}} V^{(-1)}_{w} V^{(-1/2,-1/2)}_{\bar{\mathcal{F}}(-)} 
\rangle \! \rangle, \quad 
\langle \! \langle 
V^{(0)}_{\bar{X}^{\dagger}} V^{(-1)}_{w} V^{(-1/2,-1/2)}_{\mathcal{F}(-)} 
\rangle \! \rangle.
\nonumber \\
\label{eq:non-zero_amplitudes}
\end{eqnarray}
From (\ref{eq:disk_gen}), the amplitude including $Y_m, a'_m, 
\overline{\mathcal{F}}_{(\dot{\alpha} \dot{\beta})[IJ]}$ 
takes the form of 
\begin{eqnarray}
\langle \! \langle 
V^{(0)}_Y V^{(-1)}_{a'} V^{(-1/2,-1/2)}_{\bar{\mathcal{F}}(-)} 
\rangle \! \rangle
&=& \frac{1}{2\pi^2 \alpha^{\prime 2}} \frac{1}{\kappa g_0^2} (2\pi 
\alpha')^2 g_0^2 
\left(
\frac{4 \pi^2}{\sqrt{2}}
\right)^2 \mathrm{tr}_k 
\left[
Y_m a'_n (2\pi \alpha')^{\frac{1}{2}} \overline{\mathcal{F}}_{(\dot{\alpha} \dot{\beta})[IJ]}
\right]
\nonumber \\
& & \qquad \times \int^{y_1}_{- \infty} \! d y_2 \ (y_1 - z) (y_1 - 
\bar{z}) (z - \bar{z}) \nonumber \\
& & \qquad \times \langle e^{- \phi (y_2)} e^{- \frac{1}{2} \phi (z)} 
e^{- \frac{1}{2} \phi (\bar{z})} \rangle \nonumber \\
& & \qquad \times \langle \psi^m \bar{\psi} (y_1) \psi^n (y_2) S^{\dot{\alpha}} (z) S^I (z) 
S^{\dot{\beta}} (\bar{z}) S^{J} (\bar{z}) 
\rangle.
\label{amplitude1}
\end{eqnarray}
The ghost part correlator is calculated as 
\begin{eqnarray}
\langle 
e^{- \phi (y_2)} e^{- \frac{1}{2} \phi (z)} e^{- \frac{1}{2} \phi (\bar{z})}
\rangle
= 
(y_2 - z)^{- \frac{1}{2}} (y_2 - \bar{z})^{- \frac{1}{2}} (z - 
\bar{z})^{- \frac{1}{4}}.
\label{ghost_correlator}
\end{eqnarray}
Since the spinor and the $SU(2)_I$ indices in the 
last correlator in (\ref{amplitude1}) 
are contracted with the (S,A)-type background, 
the spinor indices are symmetrized and the $SU(2)_I$ indices are 
anti-symmetrized. 
Dropping the terms that vanish after the contraction with the (S,A)-type 
background, 
the correlator becomes 
\begin{eqnarray}
& & \langle
\psi^m \bar{\psi} (y_1) \psi^n (y_2) S^{\dot{\alpha}} (z) S^I (z) 
S^{\dot{\beta}} (\bar{z}) S^{J} (\bar{z}) 
\rangle 
\nonumber \\
& & \qquad = - \epsilon^{IJ} \epsilon^{\dot{\beta} \dot{\gamma}}
(\bar{\sigma}^{mn})^{\dot{\alpha}} {}_{\dot{\gamma}} 
(y_1 - z)^{-1} (y_1 - \bar{z})^{-1} (y_2 - z)^{- \frac{1}{2}} 
(y_2 - \bar{z})^{-\frac{1}{2}} (z - \bar{z})^{\frac{1}{4}},
\label{correlator_normal}
\end{eqnarray}
where we have used 
the formula 
for the correlator including the normal ordered world-sheet 
fermions and spin fields \cite{KoLeLeSaWa}.

After evaluating these correlators, we are left with the world-sheet 
integration of the unfixed positions of vertex operators. 
We fix the three positions to $z = i, \bar{z} = -i, y_1 
\to \infty$ and there remains $y_2$ integration in the amplitudes.
It is easily calculated by using the formula
\begin{eqnarray}
\int^{\infty}_{-\infty} \! d y_2 \ \frac{1}{y_2^2 + 1} = \pi.
\label{eq:ws_int}
\end{eqnarray}
Therefore we 
obtain 
\begin{equation}
\langle \! \langle 
V^{(0)}_Y V^{(-1)}_{a'} V^{(-1/2,-1/2)}_{\bar{\mathcal{F}}(-)} 
\rangle \! \rangle
= \frac{2\pi^2}{\kappa} 
\mathrm{tr}_k 
\left[
4 \sqrt{2} \pi Y_m a'_n 
(2 \pi \alpha')^{\frac{1}{2}} 
\epsilon^{\dot{\beta} \dot{\gamma}} 
(\bar{\sigma}^{mn})^{\dot{\alpha}} {}_{\dot{\gamma}} \epsilon^{IJ} \overline{\mathcal{F}}_{(\dot{\alpha} \dot{\beta}) 
[IJ]}
\right].
\label{eq:amp1}
\end{equation}

Other amplitudes in (\ref{eq:non-zero_amplitudes}) can be 
calculated similarly,
which are given by 
\begin{eqnarray}
\langle \! \langle 
V^{(0)}_{Y^{\dagger}} V^{(-1)}_{a'} V^{(-1/2,-1/2)}_{\mathcal{F}(-)}
\rangle \! \rangle 
&=& \frac{2\pi^2}{\kappa} 
\mathrm{tr}_k 
\left[
4 \pi \sqrt{2} Y^{\dagger}_m a'_n (2\pi \alpha')^{\frac{1}{2}}
\epsilon^{\dot{\beta} \dot{\gamma}}
(\bar{\sigma}^{mn})^{\dot{\alpha}} {}_{\dot{\gamma}}
\epsilon^{I'J'} \mathcal{F}_{(\dot{\alpha} \dot{\beta}) [I'J'])}
\right], \nonumber \\
\langle \! \langle 
V^{(0)}_X V^{(-1)}_{\bar{w}} V^{(-1/2,-1/2)}_{\bar{\mathcal{F}}(-)} 
\rangle \! \rangle 
&=& \frac{2\pi^2}{\kappa} \mathrm{tr}_k 
\left[
2 \pi \sqrt{2} X^{\dot{\alpha}} \bar{w}^{\dot{\beta}} 
(2\pi \alpha')^{\frac{1}{2}} 
\epsilon^{IJ} \overline{\mathcal{F}}_{(\dot{\alpha} \dot{\beta})[IJ]}
\right], \nonumber \\
\langle \! \langle 
V^{(0)}_{X^{\dagger}} V^{(-1)}_{\bar{w}} V^{(-1/2,-1/2)}_{\mathcal{F}(-)} 
\rangle \! \rangle 
&=& \frac{2\pi^2}{\kappa} \mathrm{tr}_k 
\left[
2 \pi \sqrt{2} X^{\dagger \dot{\alpha}} \bar{w}^{\dot{\beta}} 
(2\pi \alpha')^{\frac{1}{2}} \epsilon^{I'J'} 
\mathcal{F}_{(\dot{\alpha} \dot{\beta})[I'J']}
\right], \nonumber \\
\langle \! \langle 
V^{(0)}_{\bar{X}} V^{(-1)}_{w} V^{(-1/2,-1/2)}_{\bar{\mathcal{F}}(-)} 
\rangle \! \rangle
&=& \frac{2\pi^2}{\kappa} \mathrm{tr}_k 
\left[
2 \pi \sqrt{2} w^{\dot{\alpha}} \bar{X}^{\dot{\beta}} 
(2\pi \alpha')^{\frac{1}{2}} 
\epsilon^{IJ} \overline{\mathcal{F}}_{(\dot{\alpha} \dot{\beta})[IJ]}
\right], \nonumber \\
\langle \! \langle 
V^{(0)}_{\bar{X}^{\dagger}} V^{(-1)}_{w} V^{(-1/2,-1/2)}_{\mathcal{F}(-)} 
\rangle \! \rangle
&=& \frac{2\pi^2}{\kappa} \mathrm{tr}_k 
\left[
2\pi \sqrt{2} w^{\dot{\alpha}} \bar{X}^{\dagger \dot{\beta}} 
(2\pi \alpha')^{\frac{1}{2}} \epsilon^{I'J'} \mathcal{F}_{(\dot{\alpha} \dot{\beta}) [I'J']}
\right],
\label{eq:amp2}
\end{eqnarray}
where the anti-symmetric symbols $\epsilon^{I'J'}, \epsilon_{I'J'}$ have been introduced with the 
definition $\epsilon_{43} = \epsilon^{34} = 1$. 
We leave the detailed
calculations of these amplitudes to Appendix C.

Next let us consider the amplitudes that contain the self-dual (S,A)-type background.
Because these amplitudes have been investigated in \cite{BiFrFuLe}, we show the 
results without 
explaining detail of the calculations.

The non-zero amplitudes which contain $\mathcal{F}^{(\alpha \beta)[IJ]}$ 
and $\mathcal{F}^{(\alpha \beta)[I'J']}$ are given by 
\footnote{Here, we have added a phase $i$ to 
$\mathcal{M}^{I=2}$ to make it be consistent with 
conventions in 
\cite{ItNaSa2, ItNaSaSa}.
}
\begin{eqnarray}
\langle \! \langle 
V^{(0)}_{Y^{\dagger}} V^{(-1)}_{a'} V^{(-1/2,-1/2)}_{{\mathcal{F}}(+)} 
\rangle \! \rangle 
&=&
\frac{2\pi^2}{\kappa} \mathrm{tr}_k 
\left[
4 \pi \sqrt{2} Y^{\dagger}_m a'_n 
(2\pi \alpha')^{\frac{1}{2}} \epsilon_{\beta \gamma} 
(\sigma^{mn})_{\alpha} {}^{\gamma} \epsilon_{IJ} 
\mathcal{F}^{(\alpha \beta)[IJ]}
\right],
\nonumber \\
\langle \! \langle 
V^{(0)}_Y V^{(-1)}_{a'} V^{(-1/2,-1/2)}_{\bar{\mathcal{F}}(+)} 
\rangle \! \rangle 
&=& \frac{2\pi^2}{\kappa} \mathrm{tr}_k 
\left[
4 \pi \sqrt{2} Y_m a'_n (2\pi \alpha')^{\frac{1}{2}} 
\epsilon_{\beta \gamma} 
(\sigma^{mn})_{\alpha} {}^{\gamma} \epsilon_{I'J'} \mathcal{F}^{(\alpha \beta)[I'J']}
\right], 
\nonumber \\
\langle \! \langle 
V^{(-1/2)}_{\mathcal{M}} V^{(-1/2)}_{\mathcal{M}} V^{(-1/2,-1/2)}_{\bar{\mathcal{F}}(+)} 
\rangle \! \rangle 
&=& \frac{2\pi^2}{\kappa}
\mathrm{tr}_k
\left[
- 2 \pi i \epsilon_{IJ} \mathcal{M}'_{\alpha} {}^I 
\mathcal{M}_{\beta} {}^J 
(2\pi \alpha')^{\frac{1}{2}} \epsilon_{I'J'} \overline{\mathcal{F}}^{(\alpha \beta)[I'J']}
\right].
\nonumber \\
\label{eq:amp3}
\end{eqnarray}

Finally, 
we study amplitudes that contain the (A,S)-type 
backgrounds. 
We find that the amplitudes that involve the backgrounds $\mathcal{F}^{[\alpha 
\beta](IJ)}, \mathcal{F}^{[\alpha \beta] (I'J')}, 
\mathcal{F}_{[\dot{\alpha} \dot{\beta}](I'J')}$ vanish while the 
amplitudes that contain $\mathcal{F}_{[\dot{\alpha} \dot{\beta}](IJ)}$ are non-zero.
The amplitudes including $\mathcal{F}_{[\dot{\alpha} \dot{\beta}] 
(IJ)}$ have been calculated in 
\cite{ItNaSaSa}.
The results are 
\begin{eqnarray}
\langle \! \langle
V^{(-1/2)}_{\bar{\mu}} V^{(-1/2)}_{\mu} 
V^{(-1/2,-1/2)}_{\bar{\mathcal{F}}} 
\rangle \! \rangle
&=& \frac{2\pi^2}{\kappa} 
\mathrm{tr}_k 
\left[
2 \pi i \bar{\mu}^I \mu^J (2\pi \alpha')^{\frac{1}{2}} 
\epsilon^{\dot{\alpha} \dot{\beta}}
\bar{\mathcal{F}}_{[\dot{\alpha} \dot{\beta}](IJ)}
\right], 
\nonumber \\
\langle \! \langle 
V^{(-1/2)}_{\bar{\mu}} V^{(-1/2)}_{\mu} V^{(-1/2,-1/2)}_{\bar{\mathcal{F}}} 
\rangle \! \rangle
&=& \frac{2\pi^2}{\kappa} \mathrm{tr}_k 
\left[
2 \pi i \mathcal{M}^{\prime \alpha I} \mathcal{M}'_{\alpha} {}^J 
(2\pi \alpha')^{\frac{1}{2}} \epsilon^{\dot{\alpha} \dot{\beta}}
\bar{\mathcal{F}}_{[\dot{\alpha} \dot{\beta}](IJ)}
\right].
\label{eq:amp4}
\end{eqnarray}
We note that in the D$(-1)$ effective action, the (A,S)-background 
induces mass terms of the fermionic moduli $\mu, \bar{\mu}$ and $\mathcal{M}'$.
This result can be 
also seen in the field theory 
calculations of the instanton effective action of $\mathcal{N} = 2$ 
super Yang-Mills theory realized on the D3-branes.
The (A,S)-type deformation of the spacetime actions of 
super Yang-Mills theories 
has been studied in our 
previous papers 
\cite{ItNaSa2, ItNaSaSa}, 
where the (A,S)-type background 
$\mathcal{F}_{[\dot{\alpha} \dot{\beta}] (IJ)}$ induces a chiral mass term of 
the gaugino in $\mathcal{N} = 2$ super Yang-Mills theory.
This term induces the mass terms for the fermionic moduli.

On the other hand, the background $\mathcal{F}^{[\alpha \beta] (IJ)}$ 
induces an anti-chiral mass term of the gaugino,
which 
is sub-leading order in $g$ 
and is irrelevant 
in the instanton effective action. 
The backgrounds $\mathcal{F}^{[\alpha \beta](I'J')}$ 
and $\mathcal{F}_{[\dot{\alpha} \dot{\beta}](I'J')}$ induce mass terms 
for the $\mathcal{N} = 2$ adjoint hypermultiplet fermions,
which do not appear in $\mathcal{N} = 2$ super Yang-Mills theory.

\subsection{Deformed D$(-1)$-brane effective action}
The amplitudes (\ref{eq:amp1})--(\ref{eq:amp4}) 
are reproduced by the following interaction terms in the low-energy
effective action 
of the D3/D$(-1)$-brane system 
\begin{eqnarray}
\delta \tilde{S} \!\! &=& \!\!
\frac{2\pi^2}{\kappa}
\mathrm{tr}_k 
\bigg[
2 C^{+mn} Y^{\dagger}_m a'_n 
+ 2 \bar{C}^{+mn} Y_m a'_n
- \frac{\sqrt{2}}{2} i \epsilon_{IJ} \mathcal{M}'_{\alpha} {}^I 
\mathcal{M}'_{\beta} {}^J \bar{C}^{+(\alpha \beta)}
\nonumber \\
& & \qquad \qquad 
+ 2 Y_m a'_n \bar{C}^{-mn}
+ 2 Y^{\dagger}_m a'_n C^{-mn}
+ \frac{1}{2} X_{\dot{\alpha}} \bar{w}_{\dot{\beta}} 
\epsilon^{\dot{\beta} \dot{\gamma}} (\bar{\sigma}_{mn})^{\dot{\alpha}} {}_{\dot{\gamma}} 
\bar{C}^{-mn}
\nonumber \\
& & \qquad \qquad 
+ \frac{1}{2} X^{\dagger}_{\dot{\alpha}} \bar{w}_{\dot{\beta}} 
\epsilon^{\dot{\beta} \dot{\gamma}} (\bar{\sigma}_{mn})^{\dot{\alpha}} {}_{\dot{\gamma}} C^{-mn}
+ \frac{1}{2} w_{\dot{\alpha}} \bar{X}_{\dot{\beta}}
\epsilon^{\dot{\beta} \dot{\gamma}} (\bar{\sigma}_{mn})^{\dot{\alpha}} {}_{\dot{\gamma}} 
\bar{C}^{-mn}
\nonumber \\
& & 
\qquad \qquad 
 - \frac{1}{2} w_{\dot{\alpha}} \bar{X}^{\dagger}_{\dot{\beta}} 
\epsilon^{\dot{\beta} \dot{\gamma}} (\bar{\sigma}_{mn})^{\dot{\alpha}} {}_{\dot{\gamma}} 
C^{-mn} - \mathcal{M}^{\prime \alpha I} \mathcal{M}'_{\alpha} {}^J m_{(IJ)}
- 2 \bar{\mu}^I \mu^J m_{(IJ)}
\bigg],
\end{eqnarray}
where we have defined the deformation parameters
\begin{eqnarray}
C^{+mn} &=& - 2 \pi \sqrt{2} (2 \pi \alpha')^{\frac{1}{2}} 
\epsilon_{\beta \gamma} (\sigma^{mn})_{\alpha} {}^{\gamma} 
\epsilon_{IJ} \mathcal{F}^{(\alpha \beta) [IJ]}, 
\nonumber \\
C^{-mn} &=& - 2 \pi \sqrt{2} (2 \pi \alpha')^{\frac{1}{2}} 
\epsilon^{\dot{\beta} \dot{\gamma}} (\bar{\sigma}^{mn})^{\dot{\alpha}} {}_{\dot{\gamma}} \epsilon^{I'J'} 
\mathcal{F}_{(\dot{\alpha} \dot{\beta}) 
[I'J']}, 
\nonumber \\
\bar{C}^{+mn} &=& - 2 \pi \sqrt{2} (2 \pi \alpha')^{\frac{1}{2}} 
\epsilon_{\beta \gamma} (\sigma^{mn})_{\alpha} {}^{\gamma} \epsilon_{I'J'} \overline{\mathcal{F}}^{(\alpha \beta) 
[I'J']}, 
\nonumber \\
\bar{C}^{-mn} &=& - 2 \pi \sqrt{2} (2 \pi \alpha')^{\frac{1}{2}} 
\epsilon^{\dot{\beta} \dot{\gamma}} (\bar{\sigma}^{mn})^{\dot{\alpha}} {}_{\dot{\gamma}}
 \epsilon^{IJ} \overline{\mathcal{F}}_{(\dot{\alpha} \dot{\beta}) [IJ]},
\end{eqnarray}
and the mass matrix
\begin{equation}
m_{(IJ)} = \pi i (2\pi 
\alpha')^{\frac{1}{2}} \bar{\mathcal{F}}_{[\dot{\alpha} \dot{\beta}] (IJ)} 
\epsilon^{\dot{\alpha} \dot{\beta}}. 
\end{equation}
In summary, the low-energy effective action of the D3/D$(-1)$-brane 
system in the presence of the (S,A) and (A,S)-type 
backgrounds is given by 
\begin{eqnarray}
\tilde{S}_{\mathrm{str}} (C, \bar{C}, m) 
= \tilde{S}_{\mathrm{str}} + S_{\mathrm{ADHM}} + \delta \tilde{S}.
\end{eqnarray}
After integrating out the auxiliary fields $Y, Y^{\dagger}, 
X, \bar{X}, X^{\dagger}, \bar{X}^{\dagger}$, 
we finally obtain the following 
effective action 
\begin{align}
S_{\mathrm{str}} (C, \bar{C}, m) =& \frac{2\pi^2}{\kappa} \mathrm{tr}_k 
\bigg[
- 2 
\left(
[\bar{\chi}, a'_m] + \bar{C}_{mn} a^{\prime n}
\right)
\left(
[\chi, a^{\prime m}] + C^{mk} a'_k
\right)
\notag \\
 & \qquad
+ \frac{i}{2 \sqrt{2}}
\mathcal{M}^{\prime \alpha I}
\left(
[\bar{\chi}, \mathcal{M}'_{\alpha I}]
+ \frac{1}{2} \bar{C}^{+mn} (\sigma_{mn})_{\alpha} {}^{\beta} 
\mathcal{M}'_{\beta I} + 2 \sqrt{2} i m_{(IJ)} \mathcal{M}'_{\alpha} {}^J
\right)
\notag \\
 & \qquad 
+ 
\left(\bar{\chi} \bar{w}^{\dot{\alpha}} - \bar{w}^{\dot{\alpha}} \bar{\phi}^0
+ \frac{1}{2} \bar{C}^{-mn} (\bar{\sigma}_{mn})^{\dot{\alpha}} 
{}_{\dot{\beta}} \bar{w}^{\dot{\beta}}
\right) 
\left(w_{\dot{\alpha}} \chi - \phi^0 w_{\dot{\alpha}} 
+ \frac{1}{2} C^{-mn} (\bar{\sigma}_{mn})^{\dot{\gamma}} 
{}_{\dot{\alpha}} w_{\dot{\gamma}}
\right)
\notag \\
 & \qquad
+ 
\left(
\chi \bar{w}^{\dot{\alpha}} - \bar{w}^{\dot{\alpha}} \bar{\phi}^0
+ \frac{1}{2} C^{-mn} (\bar{\sigma}_{mn})^{\dot{\alpha}} 
{}_{\dot{\beta}} \bar{w}^{\dot{\beta}}
\right) 
\left(
w_{\dot{\alpha}} \bar{\chi} - \bar{\phi}^0 w_{\dot{\alpha}} 
+ \frac{1}{2} \bar{C}^{-mn} (\bar{\sigma}_{mn})^{\dot{\gamma}} 
{}_{\dot{\alpha}} w_{\dot{\gamma}}
\right)
\notag \\
 & \qquad 
+ \frac{i}{\sqrt{2}}
\bar{\mu}^I 
\left(
- \mu_I \bar{\chi}  + \bar{\phi}^0 \mu_I 
+ 2 \sqrt{2} i \mu^J m_{(IJ)}
\right)
\notag \\
 & \qquad 
- i \bar{\psi}^{\dot{\alpha}}_I 
(\bar{\mu}^I w_{\dot{\alpha}} + \bar{w}_{\dot{\alpha}} \mu^I + 
[\mathcal{M}^{\prime \alpha I}, a'_{\alpha \dot{\alpha}}]) 
 \notag \\
 & \qquad 
+ i D^c (\tau^c)^{\dot{\alpha}} {}_{\dot{\beta}} (\bar{w}^{\dot{\beta}} 
w_{\dot{\alpha}} + \bar{a}^{\prime \dot{\beta} \alpha} a'_{\alpha \dot{\alpha}})
\bigg],
\label{eq:string_inst_action}
\end{align}
where we have defined $C_{mn} = C^{+}_{mn} + C^{-}_{mn}$, $\bar{C}_{mn} = 
\bar{C}^{+}_{mn} + \bar{C}^{-}_{mn}$.
The terms involving $C^{\pm}_{mn}, \bar{C}^{\pm}_{mn}$ in the above expression coincide 
with the parts that depend on the $\Omega$-background parameters 
$\Omega_{mn}$ and $\bar{\Omega}_{mn}$ in (\ref{insteff}) 
if we identify 
$C^{mn} = - i \Omega^{mn}, \bar{C}^{mn} = - i \bar{\Omega}^{mn}$. 

The terms that depend on the parameter $m_{(IJ)}$ 
correspond to the terms stemming from the non-zero R-symmetry Wilson line 
in (\ref{insteff}) if we choose the mass matrix as 
\begin{eqnarray}
m_{(IJ)} = - \frac{1}{2\sqrt{2}} \epsilon_{IK} \bar{\mathcal{A}}^K {}_J.
\end{eqnarray}
Therefore we conclude that the instanton effective action in the 
$\Omega$-background is obtained as the low-energy effective action of 
the D3/D$(-1)$-brane system in the presence of the constant (S,A) and (A,S)-type 
backgrounds.
In general, these backgrounds break supersymmetry.
However the choice of the background parameter 
(\ref{condition1}) ensures that the deformed 
D$(-1)$ effective action preserves one supersymmetry as discussed in section 3. 

We note that the Fayet-Iliopoulos parameter $\zeta^c$, namely the 
spacetime noncommutativity, discussed in 
section 3 is introduced as the NS-NS B-field background 
\cite{ChHo, SeWi}.

\section{Conclusions and discussion}
In this paper, 
we have studied general (non-(anti-)self-dual) $\Omega$-background deformation of the instanton effective action of $\mathcal{N}=2$ super Yang-Mills theory and the R-R 3-form background deformation of the D$(-1)$-brane effective action. We have found that these two effective actions coincide under the appropriate identification of the deformation parameters.

First we have checked the deformed supersymmetry preserved in $\mathcal{N}=2$ super Yang-Mills action in the $\Omega$-background, which was shown in \cite{NeOk}. This supersymmetry is nilpotent up to the gauge transformation by the scalar field and the rotation by the $\Omega$-background. We have found the ADHM construction of instantons in the $\Omega$-background, and derived the deformed instanton effective action of $\mathcal{N}=2$ super Yang-Mills theory.

Next we have derived the effective action of the D$(-1)$-branes in the fractional D3/D$(-1)$-brane system on the $\mathbb{C}\times\mathbb{C}^2/\mathbb{Z}_2$ orbifold in the presence of the constant R-R 3-form background by computing the disk amplitudes of open strings connecting the branes 
with an insertion of the closed string vertex operators corresponding to the R-R 3-form background. 
We have found that the D$(-1)$-brane effective action coincides with the instanton effective action in the $\Omega$-background, identifying appropriately the deformation parameters of the R-R 3-form background and the $\Omega$-background. 
Comparing the two effective actions, we have found that the (S,A)-type background deformation of the D$(-1)$-brane effective action corresponds to the deformation of the instanton effective action of $\mathcal{N}=2$ super Yang-Mills theory by the $\Omega$-background metric. 
On the other hand, the (A,S)-type background deformation of the D$(-1)$-brane effective action corresponds to the R-symmetry Wilson line in the instanton effective action.

It has been pointed out in~\cite{BiFrFuLe} that 
the self-dual part of $\Omega_{mn}$ corresponds to the constant graviphoton field strength 
in $\mathcal{N}=2$ chiral Weyl multiplet and the self-dual part of $\bar{\Omega}_{mn}$ corresponds to the field strength in $\mathcal{N}=2$ vector multiplet. 
In a recent paper~\cite{Antoniadis:2010iq}, the authors calculate
the graviphoton correction to the higher-derivative F-terms
including the anti-self-dual field strength in $\mathcal{N}=2$ vector multiplet.
In this paper we have found 
the R-R 3-form field strength which corresponds to the anti-self-dual part of $\Omega_{mn}$ and
$\bar{\Omega}_{mn}$.
It would be interesting to study the supergravity multiplet which corresponds to the anti-self-dual part of $\Omega$ and $\bar{\Omega}$.

For the application of our results to other gauge theories, it is interesting to consider $\mathcal{N}=4$ and $\mathcal{N}=2^*$ theories~\cite{Bruzzo:2002xf}.
When we put D3-branes in type IIB superstring theory without orbifold projection, we obtain $\mathcal{N}=4$ super Yang-Mills theory on the D3-branes. 
In our previous paper~\cite{ItNaSaSa} we proposed 
the extended version of $\Omega$-background for ten-dimensional metric, which induces more deformation parameters to deform four-dimensional $\mathcal{N}=4$ super Yang-Mills theory by the dimensional reduction.
The $\mathcal{N}=2^*$ theory in the ten-dimensional $\Omega$-background can be similarly obtained.
In calculating the instanton partition function of gauge theories 
using the deformation of the theories and the localization formula, 
it is one of the important steps to confirm whether the deformed theories possess some nilpotent operator.
In this paper we have obtained the supersymmetry preserved by the deformed $\mathcal{N}=2$ super Yang-Mills action, 
then it is natural to search for the similar result in $\mathcal{N}=4$ and $\mathcal{N}=2^*$ theories.

The string construction of instantons can be applied to higher dimensional instantons.
In \cite{Jafferis:2007sg} the D6/D0-brane system is considered to obtain the partition function of six-dimensional instantons.
In \cite{BiFeFrGaLePe, FuMoPo, Billo:2010mg} the authors discussed the R-R 3-form
background deformation of D7/D3/D$(-1)$-brane system to investigate the
heterotic-type I$'$ duality. 
The $\Omega$-background approach is applied to the case of M-theory
\cite{Ne-M}. 
It is interesting to study the
role of R-R backgrounds in higher dimensional gauge theories.

\subsection*{Acknowledgments}
The work of H.~N. is supported by Mid-career Researcher Program 
through the National Research Foundation of Korea(NRF) grant 
funded by the Korea government(MEST)(No. 2009-0084601). 
The work of T.~S. is supported by the Global
Center of Excellence Program by MEXT, Japan through the
"Nanoscience and Quantum Physics" Project of the Tokyo
Institute of Technology, and by Iwanami Fujukai Foundation.
The work of S.~S. is supported by the Japan Society for the Promotion of Science (JSPS) Research Fellowship.

\begin{appendix}

\section{Twisted supersymmetry transformation of the ADHM moduli}
\label{rewrite}
In the twisted supersymmetry transformation \eqref{defBRST2} 
of the ADHM moduli, 
we define new variables 
\begin{align}
B_{1}&=-ia'_{1}-a'_{2},& \psi_{B_{1}}&=-i\mathcal{M}'_{1}-\mathcal{M}'_{2},
\notag\\ 
B_{2}&=-ia'_{3}+a'_{4},& \psi_{B_{2}}&=-i\mathcal{M}'_{3}+\mathcal{M}'_{4},
\notag\\
I&=\bar{w}^{\dot{1}},& \psi^{}_{I}&=\bar{\mu}^{\dot{1}},
\notag\\
J&=w_{\dot{2}},& \psi^{}_{J}&=\mu_{\dot{2}},
\notag\\
H^{}_{\mathbb{R}}&=D_{12},& \chi^{}_{\mathbb{R}}&=\bar{\psi}_{12}, 
\notag\\
H^{}_{\mathbb{C}}&=D_{23}+iD_{31}, & 
\chi^{}_{\mathbb{C}}&=\bar{\psi}_{23}+i\bar{\psi}_{31}. 
\end{align}
We also redefine the variables as 
\begin{gather}
-2\sqrt{2}i\chi\to\phi-\frac{1}{2}(\epsilon_{1}+\epsilon_{2}),\quad 
-2\sqrt{2}i\phi^{0}\to a,\quad 
2\sqrt{2}i\bar{\chi}\to\bar{\phi},\quad 
2\sqrt{2}i\bar{\eta}\to\eta. 
\end{gather}
Then \eqref{defBRST2} is rewritten in terms of the new variables 
using \eqref{omega} as 
\begin{align}
\bar{Q}^{}_{\Omega}B_{i}&=\psi_{B_{i}}, & 
\bar{Q}^{}_{\Omega}\psi_{B_{i}}&=[\phi,B_{i}]+\epsilon_{i}B_{i}, 
\quad(i=1,2)
\notag\\
\bar{Q}^{}_{\Omega}I&=\psi^{}_{I}, & 
\bar{Q}^{}_{\Omega}\psi^{}_{I}&=\phi I-Ia,
\notag\\
\bar{Q}^{}_{\Omega}J&=\psi^{}_{J}, & 
\bar{Q}^{}_{\Omega}\psi^{}_{J}&=
-J\phi+aJ+(\epsilon_{1}+\epsilon_{2})J,
\notag\\
\bar{Q}^{}_{\Omega}\phi&=0, &&
\notag\\
\bar{Q}^{}_{\Omega}\bar{\phi}&=\eta, & 
\bar{Q}^{}_{\Omega}\eta&=[\phi,\bar{\phi}],
\notag\\
\bar{Q}^{}_{\Omega}\chi^{}_{\mathbb{R}}&=H^{}_{\mathbb{R}}, & 
\bar{Q}^{}_{\Omega}H^{}_{\mathbb{R}}&=[\phi,\chi^{}_{\mathbb{R}}],
\notag\\
\bar{Q}^{}_{\Omega}\chi^{}_{\mathbb{C}}&=H^{}_{\mathbb{C}}, & 
\bar{Q}^{}_{\Omega}H^{}_{\mathbb{C}}&=
[\phi,\chi^{}_{\mathbb{C}}]+(\epsilon_{1}+\epsilon_{2})\chi^{}_{\mathbb{C}}. 
\end{align}
This is the same as the twisted supersymmetry transformation in \cite{Ne}.

\section{World-sheet fields and orbifold projection}
In this appendix, we summarize our notations and conventions of 
world-sheet fields in type IIB superstring theory in flat 
ten-dimensional Euclidean spacetime. 
The world-sheet coordinates are denoted by $z, \bar{z}$. 
The world-sheet fields $X^M (z, \bar{z})$, $\psi^M (z), \tilde{\psi}^M 
(\bar{z}), \ (M=1, \cdots, 10)$ are bosonic and fermionic string 
coordinates. The left moving parts of the 
fields satisfy the free field OPEs given by 
\begin{eqnarray}
X^M (z) X^N (w) \sim - \delta^{MN} \ln (z - w), \qquad 
\psi^M (z) \psi^N (w) \sim \delta^{MN} (z - w)^{-1}.
\end{eqnarray}
The right moving parts satisfy the same OPEs. 
In the presence of D3-branes, the $SO(10)$ Lorentz symmetry is broken 
down to $SO(4) \times SO(6)$ and the string coordinates are decomposed 
as $X^M = (X^m, X^a), \quad \psi^M = (\psi^m, \psi^a) $, $(m=1,\cdots,4, 
a=5,\cdots,10)$, where 
$X^m$ spans the world-volume of the D3-branes and 
$X^a$ represents the transverse directions.

The spin fields in the vertex operators are defined by 
\begin{eqnarray}
\begin{aligned}
 & S^{\lambda} = e^{\lambda \phi \cdot e} c_{\lambda}, \quad 
 \phi \cdot e = \phi^i e_i \quad (i=1, \cdots, 5), \\
 & \lambda = \frac{1}{2} (\pm e_1 \pm 
e_2 \pm e_3 \pm e_4 \pm e_5) \equiv \lambda_i e_i,
\quad \lambda_i = \pm \frac{1}{2}, 
\end{aligned}
\label{spin_field}
\end{eqnarray}
where $e_i$ are the basis of $SO(10)$ spinor representation, 
$c_{\lambda}$ is the cocycle factor and $\phi^i$ are the free bosons 
for the bosonization of $\psi^M$ 
\cite{KoLeLeSaWa}. 
The weight vector $\lambda$ specifies the ten-dimensional spinor index $\hat{\mathcal{A}}$. 
There are odd numbers of minus components in the weight vector in type 
IIB string theory. The ten-dimensional 
spin fields $S^{\hat{\mathcal{A}}} \ (\hat{\mathcal{A}} = 1, \cdots, 16)$ in the 
presence of the D3-branes are decomposed as 
\begin{eqnarray}
S^{\hat{\mathcal{A}}} \to (S_{\alpha} S_A, S^{\dot{\alpha}} S^A),
\end{eqnarray}
where $\alpha, \dot{\alpha} = 1,2$ are $SO(4)$ and 
$A=1,2,3,4$ are 
$SO(6)$ spinor ($SU(4)$ vector) indices. 
Let us define the four-dimensional and the internal 
spin states which specify the weight vector $\lambda$: 
\begin{eqnarray}
| \lambda_1, \lambda_2 \rangle, \quad 
| \lambda_3, \lambda_4, \lambda_5 \rangle. 
\end{eqnarray}
The relation between the four-dimensional spinor indices and the spin states 
is found in Table \ref{4d_spin} while the 
relation between the internal spin indices and the internal spin states 
is found in Table \ref{6d_spin}.
\begin{table}[t]
\begin{center}
\begin{tabular}{|l||l|l|l|l|}
\hline
     & $\alpha = 1$ & $\alpha = 2$ & $ \dot{\alpha} = 1$ & $\dot{\alpha} 
     = 2$ \\
\hline
lower indices 
& $\left| \left. + \frac{1}{2}, + \frac{1}{2} \right\rangle \right. $ 
& $\left| \left. - \frac{1}{2}, - \frac{1}{2} \right\rangle \right. $
& $\left| \left. - \frac{1}{2}, + \frac{1}{2} \right\rangle \right. $
& $\left| \left. + \frac{1}{2}, - \frac{1}{2} \right\rangle \right. $
\\
\hline
upper indices 
& $\left| \left. - \frac{1}{2}, - \frac{1}{2} \right\rangle \right. $
& $\left| \left. + \frac{1}{2}, + \frac{1}{2} \right\rangle \right. $
& $\left| \left. + \frac{1}{2}, - \frac{1}{2} \right\rangle \right. $
& $\left| \left. - \frac{1}{2}, + \frac{1}{2} \right\rangle \right. $
\\
\hline
\end{tabular}
\end{center}
\caption{Spin states for the four-dimensional spinor indices.}
\label{4d_spin}
\end{table}
\begin{table}[t]
\begin{center}
\begin{tabular}{|l||l|l|l|l|}
\hline
     & $A = 1$ & $A = 2$ & $A = 3$ & $A = 4$ \\
\hline
lower indices 
& $\left| \left. - \frac{1}{2}, - \frac{1}{2}, - \frac{1}{2} \right\rangle \right. $ 
& $\left| \left. - \frac{1}{2}, + \frac{1}{2}, + \frac{1}{2} \right\rangle \right. $
& $\left| \left. + \frac{1}{2}, - \frac{1}{2}, + \frac{1}{2} \right\rangle \right. $
& $\left| \left. + \frac{1}{2}, + \frac{1}{2}, - \frac{1}{2} \right\rangle \right. $
\\
\hline
upper indices 
& $\left| \left. + \frac{1}{2}, + \frac{1}{2}, + \frac{1}{2} \right\rangle \right. $
& $\left| \left. + \frac{1}{2}, - \frac{1}{2}, - \frac{1}{2} \right\rangle \right. $
& $\left| \left. - \frac{1}{2}, + \frac{1}{2}, - \frac{1}{2} \right\rangle \right. $
& $\left| \left. - \frac{1}{2}, - \frac{1}{2}, + \frac{1}{2} \right\rangle \right. $
\\
\hline
\end{tabular}
\end{center}
\caption{Spin states for the internal indices.}
\label{6d_spin}
\end{table}

Now we consider the case that the internal space is orbifolded
as $\mathbb{C} \times \mathbb{C}^2/\mathbb{Z}_2$. 
The string coordinates in the transverse directions to 
the D3-branes are combined as
\begin{eqnarray}
\begin{aligned}
 & Z = \frac{1}{\sqrt{2}} (X^5 - i X^6), \qquad 
Z^{(1)} = \frac{1}{\sqrt{2}} (X^7 - i X^8), \qquad 
Z^{(2)} = \frac{1}{\sqrt{2}} (X^9 - i X^{10}), \\
 & \psi = \frac{1}{\sqrt{2}} (\psi^5 - i \psi^6), \qquad 
\psi^{(1)} = \frac{1}{\sqrt{2}} (\psi^7 - i \psi^8), \qquad 
\psi^{(2)} = \frac{1}{\sqrt{2}} (\psi^9 - i \psi^{10}).
\end{aligned}
\end{eqnarray}
The $\mathbb{Z}_2$ action is defined by $\pi$ rotation in the 7-8 plane and 
$- \pi$ rotation in the 9-10 plane. This action is given by  
\begin{eqnarray}
(Z, Z^{(1)}, Z^{(2)}) \to (Z, - Z^{(1)}, - Z^{(2)}), \qquad 
(\psi, \psi^{(1)}, \psi^{(2)}) \to (\psi, - \psi^{(1)}, - \psi^{(2)}).
\end{eqnarray}

The $SU(4)$ internal symmetry is broken to $SU(2) \times SU(2)$ 
by the orbifold projection 
and the $SU(4)$ index $A=1,2,3,4$ is decomposed into the $SU(2) \times 
SU(2)$ indices $I=1,2$ and $I' = 3,4$. 
The $\mathbb{Z}_2$ action on the spin states is expressed by 
the operator $g = \mathbf{1} \otimes i \sigma^3 \otimes (-i \sigma^3)$ acting on the 
internal spin states. The $g$-invariant states are given by 
\begin{eqnarray}
\left| \left. \frac{s}{2}, + \frac{1}{2}, + \frac{1}{2} \right\rangle 
\right., \qquad 
\left| \left. \frac{s}{2}, - \frac{1}{2}, - \frac{1}{2} \right\rangle 
\right., \qquad (s = \pm 1).
\end{eqnarray}
Therefore the $\mathbb{Z}_2$-invariant internal spin fields are given by 
$S_I$ and $S^I$.

On the other hand, the states with index $I'$ get an extra minus sign by the 
$\mathbb{Z}_2$ action.
We note that although each $S_{I'}$ is not invariant under 
the $\mathbb{Z}_2$ orbifold action, the bi-spinors, 
for example, $S_{I'} (z) S_{J'} (\bar{z})$ is invariant.

\section{Detailed calculations of the amplitudes}
In this appendix, we show the detailed calculations of the amplitudes 
(\ref{eq:amp2}) in section 4. 

The first amplitude in (\ref{eq:amp2}) is given by 
\begin{eqnarray}
\langle \! \langle 
V^{(0)}_{Y^{\dagger}} V^{(-1)}_{a'} V^{(-1/2,-1/2)}_{\mathcal{F}(-)}
\rangle \! \rangle 
&=& \frac{1}{2\pi^2 \alpha^{\prime 2}} \frac{1}{\kappa g_0^2} (2\pi 
\alpha')^2 g_0^2 
\left(
\frac{4 \pi^2}{\sqrt{2}}
\right)^2 \mathrm{tr}_k 
\left[
Y^{\dagger}_m a'_n (2\pi \alpha')^{\frac{1}{2}} \mathcal{F}_{(\dot{\alpha} \dot{\beta})[I'J']}
\right]
\nonumber \\
& & \times \int^{y_1}_{- \infty} \! d y_2 \ (y_1 - z) (y_1 - 
\bar{z}) (z - \bar{z}) 
\langle 
e^{- \phi (y_2)} e^{- \frac{1}{2} \phi (z)} 
e^{- \frac{1}{2} \phi (\bar{z})} 
\rangle \nonumber \\
& & \times 
\langle 
\psi^m \psi (y_1) \psi^n (y_2) S^{\dot{\alpha}} (z) S^{I'} (z) 
S^{\dot{\beta}} (\bar{z}) S^{J'} (\bar{z}) 
\rangle.
\end{eqnarray}
The ghost part correlator is evaluated 
in (\ref{ghost_correlator}). 
The spin field part is given by 
\begin{eqnarray}
& & \langle 
\psi^m \psi (y_1) \psi^n (y_2) S^{\dot{\alpha}} (z) S^{I'} (z) 
S^{\dot{\beta}} (\bar{z}) S^{J'} (\bar{z}) 
\rangle \nonumber \\
& & \qquad = 
- \epsilon^{I'J'} \epsilon^{\dot{\beta} \dot{\gamma}} 
(\bar{\sigma}^{mn})^{\dot{\alpha}} {}_{\dot{\gamma}}
(y_1 - z)^{-1} (y_1 - \bar{z})^{-1} (y_2 - z)^{- \frac{1}{2}} 
(y_2 - \bar{z})^{-\frac{1}{2}} (z - \bar{z})^{\frac{1}{4}},
\end{eqnarray}
where we have 
used the fact that the $SO(4)$ spinor and internal indices  
are contracted with the (S,A)-type background 
as discussed in section 4.
Using (\ref{eq:ws_int}), we find 
\begin{equation}
\langle \! \langle 
V^{(0)}_{Y^{\dagger}} V^{(-1)}_{a'} V^{(-1/2,-1/2)}_{\mathcal{F}(-)}
\rangle \! \rangle 
= \frac{2\pi^2}{\kappa} 
\mathrm{tr}_k 
\left[
4 \sqrt{2} \pi Y^{\dagger}_m a'_n (2\pi \alpha')^{\frac{1}{2}} 
\epsilon^{\dot{\beta} \dot{\gamma}} 
(\bar{\sigma}^{mn})^{\dot{\alpha}} {}_{\dot{\gamma}} \epsilon^{I'J'} 
\mathcal{F}_{(\dot{\alpha} \dot{\beta})[I'J']}
\right].
\end{equation}

The second amplitude in (\ref{eq:amp2}) is 
\begin{eqnarray}
\langle \! \langle 
V^{(0)}_X V^{(-1)}_{\bar{w}} V^{(-1/2,-1/2)}_{\overline{\mathcal{F}}(-)} 
\rangle \! \rangle
&=& \frac{1}{2\pi^2 \alpha^{\prime 2}} \frac{1}{\kappa g_0^2} (2\pi 
\alpha')^2 g_0^2 
(\sqrt{2}\pi^2) \mathrm{tr}_k 
\left[
X_{\dot{\alpha}} \bar{w}_{\dot{\beta}} (2\pi \alpha')^{\frac{1}{2}} \overline{\mathcal{F}}_{(\dot{\gamma} \dot{\delta})[IJ]}
\right]
\nonumber \\
& & \times \int^{y_1}_{- \infty} \! d y_2 \ (y_1 - z) (y_1 - 
\bar{z}) (z - \bar{z}) 
\langle e^{- \phi (y_2)} e^{- \frac{1}{2} \phi (z)} 
e^{- \frac{1}{2} \phi (\bar{z})} \rangle
\nonumber \\
& & \times 
\langle \Delta (y_1) \bar{\Delta} (y_2) \rangle 
\langle 
S^{\dot{\alpha}} (y_1) S^{\dot{\beta}} (y_2) S^{\dot{\gamma}} (z) 
S^{\dot{\delta}} (\bar{z}) \rangle \langle \bar{\psi} (y_1) S^{I} (z) 
S^{J} (\bar{z})
\rangle.
\nonumber \\
\end{eqnarray}
The ghost part correlator is evaluated as before.
The twist field part is evaluated as in 
\cite{Twist}. The result is 
\begin{eqnarray}
\langle \Delta (y_1) \bar{\Delta} (y_2) \rangle = 
(y_1 - y_2)^{-\frac{1}{2}}.
\end{eqnarray}
The correlator that includes four-dimensional spin fields
is given by 
\begin{eqnarray}
& & \langle 
S^{\dot{\alpha}} (y_1) S^{\dot{\beta}} (y_2) S^{\dot{\gamma}} (z) 
S^{\dot{\delta}} (\bar{z}) 
\rangle \nonumber \\
& & = 
\epsilon^{\dot{\alpha} \dot{\delta}} \epsilon^{\dot{\beta} \dot{\gamma}} 
(y_1 - y_2) (z - \bar{z})
\left[
(y_1 - y_2) (y_1 - z) (y_1 - \bar{z})
(y_2 - z) (y_2 - \bar{z}) (z - \bar{z})
\right]^{-\frac{1}{2}},
\nonumber \\
\end{eqnarray}
where we have 
used the fact that the spin indices are contracted with the 
(S,A)-type background.
The internal space part is given by 
\begin{eqnarray}
\langle 
\bar{\psi} (y_1) S^{I} (z) S^{J} (\bar{z})
\rangle
= \epsilon^{IJ} (y_1 - z)^{- \frac{1}{2}} (y_1 - \bar{z})^{- \frac{1}{2}} 
(z - \bar{z})^{- \frac{1}{4}}.
\end{eqnarray}
Using (\ref{eq:ws_int}), we obtain 
\begin{eqnarray}
\langle \! \langle 
V^{(0)}_X V^{(-1)}_{\bar{w}} V^{(-1/2,-1/2)}_{\bar{\mathcal{F}}(-)} 
\rangle \! \rangle 
&=& \frac{2\pi^2}{\kappa} \mathrm{tr}_k 
\left[
2 \pi \sqrt{2} X^{\dot{\alpha}} \bar{w}^{\dot{\beta}} (2\pi 
\alpha')^{\frac{1}{2}} \epsilon^{IJ} \bar{\mathcal{F}}_{(\dot{\alpha} \dot{\beta})[IJ]}
\right].
\end{eqnarray}

The third amplitude in (\ref{eq:amp2}) is 
\begin{eqnarray}
\langle \! \langle 
V^{(0)}_{X^{\dagger}} V^{(-1)}_{\bar{w}} V^{(-1/2,-1/2)}_{\mathcal{F}(-)} 
\rangle \! \rangle
&=& \frac{1}{2\pi^2 \alpha^{\prime 2}} \frac{1}{\kappa g_0^2} (2\pi 
\alpha')^2 g_0^2 
(\sqrt{2}\pi^2) \mathrm{tr}_k 
\left[
X^{\dagger}_{\dot{\alpha}} \bar{w}_{\dot{\beta}} (2\pi \alpha')^{\frac{1}{2}} 
\mathcal{F}_{(\dot{\gamma} \dot{\delta})[I'J']}
\right]
\nonumber \\
& & \times \int^{y_1}_{- \infty} \! d y_2 \ (y_1 - z) (y_1 - 
\bar{z}) (z - \bar{z}) 
\langle e^{- \phi (y_2)} e^{- \frac{1}{2} \phi (z)} 
e^{- \frac{1}{2} \phi (\bar{z})} \rangle
\nonumber \\
& & \times 
\langle 
\Delta (y_1) \bar{\Delta} (y_2) 
\rangle 
\langle 
S^{\dot{\alpha}} (y_1) S^{\dot{\beta}} (y_2) S^{\dot{\gamma}} (z) 
S^{\dot{\delta}} (\bar{z}) 
\rangle 
\langle 
\psi (y_1) S^{I'} (z) S^{J'} (\bar{z})
\rangle.
\nonumber \\
\end{eqnarray}
The correlators with the ghost, the twist fields and the four-dimensional spin fields are evaluated 
as before. The internal spin field part becomes
\begin{eqnarray}
\langle 
\psi (y_1) S^{I'} (z) S^{J'} (\bar{z})
\rangle = \epsilon^{I'J'} (y_1 - z)^{- \frac{1}{2}} (y_1 - \bar{z})^{- 
\frac{1}{2}} (z - \bar{z})^{- \frac{1}{4}}.
\end{eqnarray}
Using (\ref{eq:ws_int}), we obtain
\begin{eqnarray}
\langle \! \langle 
V^{(0)}_{X^{\dagger}} V^{(-1)}_{\bar{w}} V^{(-1/2,-1/2)}_{\mathcal{F}(-)} 
\rangle \! \rangle
&=& \frac{2\pi^2}{\kappa} \mathrm{tr}_k 
\left[
2 \pi \sqrt{2} X^{\dagger \dot{\alpha}} \bar{w}^{\dot{\beta}} (2\pi 
\alpha')^{\frac{1}{2}} \epsilon^{I'J'} \mathcal{F}_{(\dot{\alpha} \dot{\beta})[I'J']}
\right].
\end{eqnarray}
The last two amplitudes in (\ref{eq:amp2}) 
are evaluated similarly. The world-sheet correlators in the amplitudes 
are calculated as before.

\end{appendix}

\end{document}